\documentclass{aa}
\usepackage{graphicx}
\usepackage{psfrag}
\def\lsim{\raise0.3ex\hbox{$<$}\kern-0.75em{\lower0.65ex\hbox{$\sim$}}}
\def\gsim{\raise0.3ex\hbox{$>$}\kern-0.75em{\lower0.65ex\hbox{$\sim$}}}                 
\begin{document}

\title{Measuring the properties of extragalactic dust and
implications for the Hubble diagram}

\author{
Ariel Goobar, Lars Bergstr\"om \& Edvard M\"ortsell}
\institute{
  Department of Physics, Stockholm University, SCFAB, S-106 91 Stockholm, Sweden. \\
  e-mail:ariel@physto.se,lbe@physto.se,edvard@physto.se
}
\authorrunning{Goobar, Bergstr\"om \& M\"ortsell}

\titlerunning{Measuring the properties of extragalactic dust}

\offprints{A. Goobar}

\date{Received ...; accepted ...}

\abstract{
Scattering and absorption of light 
by a homogeneous distribution of intergalactic large dust grains has been 
proposed as an alternative, non-cosmological explanation for the faintness of Type Ia supernovae at $z\sim 0.5$. 
We investigate the differential extinction for high-redshift
sources caused by extragalactic dust along the line of sight.

Future observations of Type Ia supernovae up to $z\sim 2$, e.g. by 
the proposed SNAP satellite, will allow the measurement of the properties of dust 
over cosmological distances. We show that 1\% {\em relative} spectrophotometric accuracy 
(or broadband photometry) in 
the wavelength interval 0.7--1.5 $\mu$m is required to measure the  
extinction caused by ``grey'' dust down to $\delta m=0.02$ magnitudes.  

We also argue that the presence of grey dust is not necessarily inconsistent with 
the recent measurement of the brightness of a supernova at $z=1.7$ (SN 1997ff),
in the absence of  accurate spectrophotometric information of the supernova.

}

\maketitle
\section{Introduction}

There is observational evidence (\cite{scp,highz}) 
that Type Ia supernovae, when used as calibrated 
standard candles, are dimmer at high redshift than can be explained
in models without a cosmological constant.  At least there seems 
to be a need for a
non-clustered ``dark energy'' 
component with a negative coefficient in the equation
of state, such as obtained in models with an evolving scalar
field (``quintessence'') (\cite{steinhardt:haga}). Since there are potential systematic 
effects affecting this interpretation, it is
important to investigate alternative explanations.
In this note, we investigate dimming due to absorption and scattering
on intergalactic dust, as has been proposed by Aguirre (\cite{aguirre99a,aguirre99b}) to be a viable
explanation for the supernova results. The recent measurements of
the CMB small angle anisotropies showing that the universe is most likely flat
(\cite{boomerang,maxima,dasi}), combined with measurements of
$\Omega_{\rm M}\sim 0.3$ from large scale structure (\cite{2DF})
and galaxy cluster evolution (\cite{bahcall}) makes Aguirre's idea for the origin of the 
Type Ia supernovae faintness at $z\sim 0.5$ unlikely. 

Moreover, \cite{aguirrehaiman} have shown
that the amount of dust required to make the supernova results compatible with 
a flat universe, as indicated by the CMB results, but {\em without} a cosmological constant, is
already disfavored by the far-infrared background measured by the DIRBE/FIRAS instruments.
 However, a sizable grey dust
column density capable of biasing the results cannot be excluded with
the present knowledge. Precision measurements of ``Dark Energy'' and ``Dark Matter'' with high-z supernovae have
been proposed see e.g. (\cite{goliath,mortsell}). The assumption in those studies is that the  systematic uncertainties
do not exceed $\delta m$=0.02 mag. Thus, in this work we concentrate on the needed relative 
spectrophotometric accuracy to meet this requirement.

Extinction  must be considered  for
at least four different dust environments:
\begin{enumerate}
\item A homogeneous intergalactic dust component. Particular attention
must be paid to the possibility that the mechanisms expelling dust
from galaxies and clusters destroy the smallest grains, thereby
causing very little reddening (``grey dust'').
\item A host galaxy dust component.
\item Dust in galaxies between the source and the observer. 
\item Milky-Way dust
\end{enumerate}

Properties such as the extinction scale-length and the wavelength 
dependence can be different
for each case. 

The main emphasis of this work is on the calculation of the possible effects from 
intergalactic dust (1). To avoid
observational constraints on reddening, such a component must
mainly consist of 
large dust grains as
described in  (Aguirre \cite{aguirre99a,aguirre99b}). However, even large-grain
dust will
cause {\em some} reddening, and going to higher redshift should enable
to distinguish between extinction or a cosmological origin  for the 
faintness of  Type Ia supernovae at z$\sim$0.5.
The future data sample
we have in mind here is the one expected from the SNAP satellite
(\cite{snap}), which will provide several thousand Type Ia supernovae
out to $z\sim 1.7$.

In this note we also discuss the light extinction by ``normal'' dust in either 
the SN host galaxy (2) or galaxies along the line of sight (3). As
we will see, the latter
only affects sizably about 1 \% of the supernovae at $z\sim 1$.

\section{Spectrophotometric calibrators}

\begin{figure}
 \centering
 \includegraphics[width=\hsize]{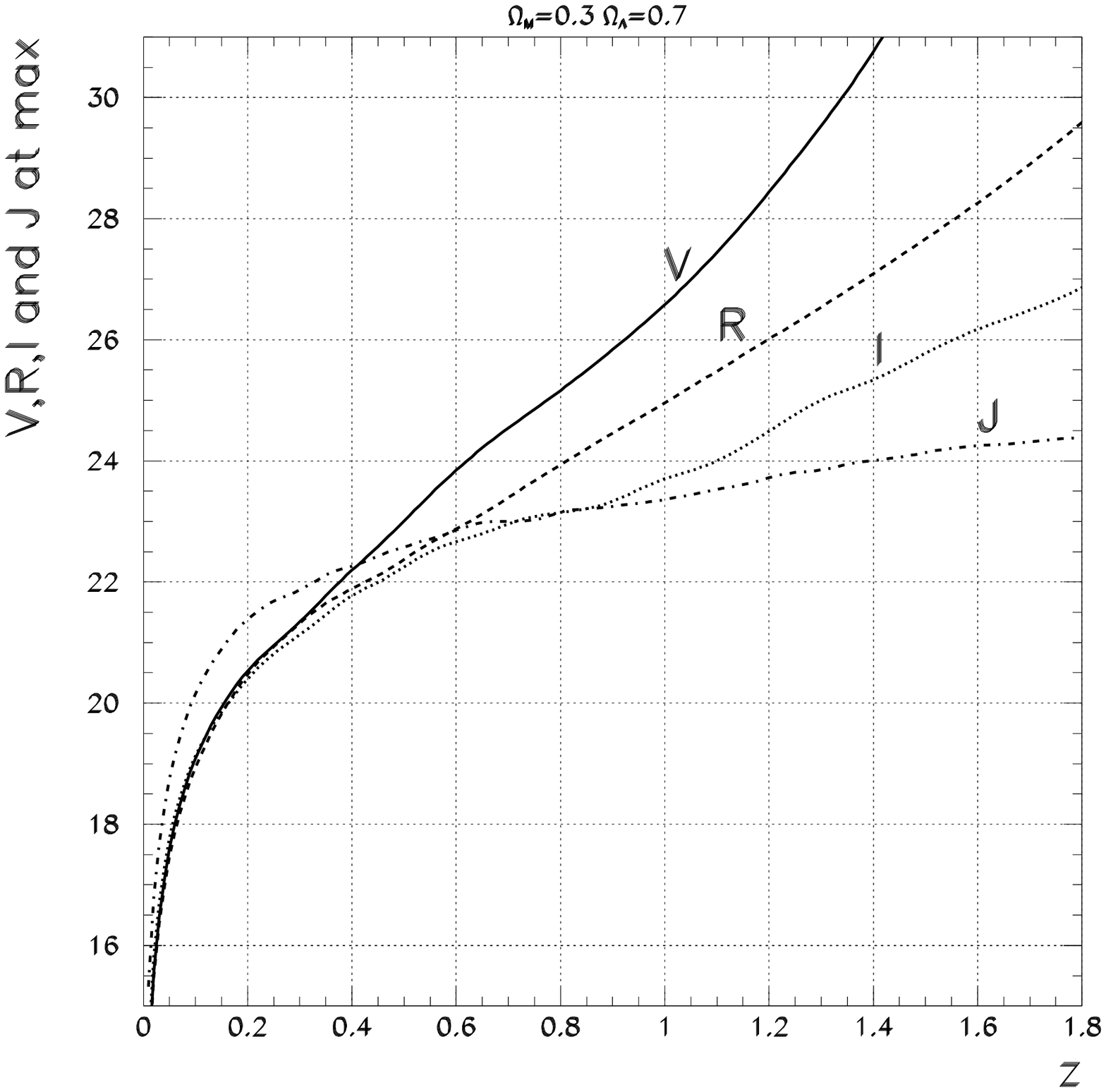} 
 \includegraphics[width=\hsize]{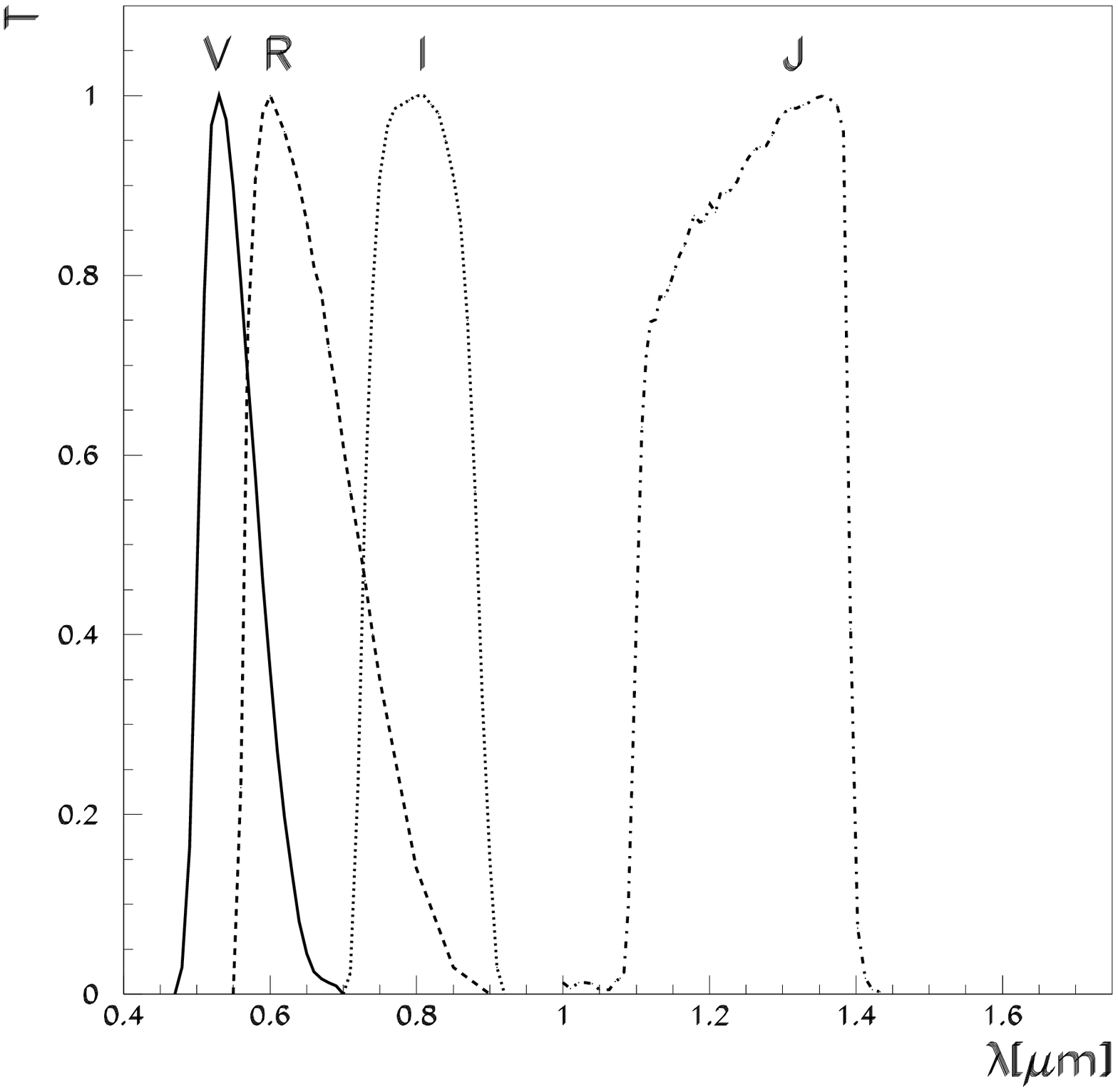} 
 \caption{Top: mean VRIJ magnitudes for Type Ia supernovae vs redshift for
             ($\Omega_{\rm M}$,$\Omega_\Lambda$)=(0.7,0.3). Bottom: filter transmission functions used.}
 \label{fig:magvsz} 
\end{figure}

Type Ia supernovae form a remarkably homogeneous class of astronomical
objects and are therefore well suited for studying the dust properties
over cosmological distances.  Figure \ref{fig:magvsz} shows the average
brightness at maximum light in V,R, I and J-band for Type Ia
supernovae along with the filter transmission functions
used.\footnote{Spectral template kindly provided by P.Nugent
(\cite{nugent}).}  The magnitudes were calculated for a ``standard'',
$\Lambda$-dominated universe, ($\Omega_{\rm M}$,$\Omega_\Lambda$)=(0.3,0.7),
with no dust extinction included.  The flux in different
band-passes is dust model and cosmology dependent as the total
extinction at any wavelength depends on the total dust column depth. 
This, in turn, depends on cosmological
parameters and the assumed density of dust to explain the faintness of supernovae
at $z\sim 0.5$. To date, one supernova at $z=0.46$ for
which there exists NIR data has been used to place limits on grey dust
(\cite{riessNIR}).
%\hspace*{-0.7cm} 

Other potential sources for spectrophotometric studies of dust are
quasars and core collapse supernovae.  The SDSS and 2DF
collaborations are preparing templates that will eventually include
about 10$^5$ QSOs up to redshift $z\sim 5$.  A recent study
 performed by the SDSS group of a sample of 2200
QSO spectra in the redshift range $0.044<z<4.789$ showed a
spectrum-to-spectrum 1 $\sigma$ difference of approximately 20\% in
the rest system wavelength range 0.1 to 0.85 $\mu$m (\cite{SDSSquasars}). By averaging over
some $400$ QSOs in redshift bins of $\Delta z=0.1$, one could thus
possibly achieve a 1\% spectrophotometric measurement, if the scatter
is of stochastic nature, i.e. not related to evolutionary effects.
Type II supernovae exhibit an approximate black body spectrum and are
used for distance estimations through the ``Expanding Photosphere
Method''. Although fainter than Type Ia and thus harder to observe at
high-z, they could potentially be used for probing dust, especially if
observed with NGST.

\section{Large ``grey''  dust grains}
Following the idea of Aguirre
(\cite{aguirre99a,aguirre99b}), we assume that the intergalactic
dust population can be described by a Draine \& Lee model (\cite{drainelee})
where the smaller grains have been destroyed by some unspecified process,
plausibly connected to the expulsion of the dust from the star-forming 
galaxies where the dust originated.

The reddening parameter $R_V$ is defined by
$$ A_V=R_VE(B-V),$$
where $A_V$ is the wavelength-dependent extinction and
$$ E(B-V)=(B-V)-(B-V)_i$$ 
with $(B-V)_i$ being the intrinsic (unobscured) color.

The presence of small dust grains enhances the wavelength dependence
of extinction. In the adopted parameterization this corresponds to lower
values of $R_V$. Measurements of the restframe E(B-V)
for $z \sim 0.5$ supernovae have been used to set a lower limit $R_V \ \gsim \ 6$ (Perlmutter et al. 1999), 
see also caveats with regards to this limit in (Aguirre 1999b). 

A population of large grains may have an $R_V$ parameter as
large as 5 to 10, thus giving a rather achromatic (``grey'') absorption.

Typically, the most important types of intergalactic grains are silicates
and graphite. The optical properties, of which $R_V$ is the most
important for our applications, depend to some extent on the actual
value chosen for the small-size cutoff $a_{\rm min}$ 
in the Draine-Lee power-law
size distribution. To make contact with Aguirre's calculations, we will
use $a_{\rm min}$ between 0.08 and 0.12 $\mu$m, corresponding
roughly to $R_V$ between 5.5 and 9.5. 
In the numerical calculations, we use the convenient parameterization
of the extinction versus wavelength given by Cardelli, Clayton \& Mathis (1989).
The existence of small dust grains would make the differential extinction effects
larger than the ones described here, i.e. easier to identify observationally.

%\subsection{Simulations}

\subsection{Extinction by dust at cosmological distances} 
For a  given emission redshift $z_e$, the attenuation $\Delta m_{\rm d}$ at observed
wavelength $\lambda_o$ 
due to dust can be written 

\begin{eqnarray}
\Delta m_{\rm d}(z_e,\lambda_o)=
&&\nonumber\\
-2.5\log_{10}\left[e^{-C\int_0^{z_e}\rho_{\rm dust}(z)a(\lambda_o/(1+z),R_V)h(z)dz}\right],\label{eq:deltam}
\end{eqnarray}
where $\rho_{\rm dust}(z)$ is the physical dust density at redshift $z$, $a(\lambda,R_V)$
is the wavelength-dependent attenuation (``reddening'') parametrized as in
(\cite{ccm}), and the cosmology-dependent function $h(z)$ is given by

\begin{equation}
h(z)={1\over H_0\left(1+z\right)\sqrt{(1+z)^2(1+\Omega_{\rm M}z)-z(2+z)\Omega_\Lambda}   }.
\end{equation}

The normalization constant $C$ is related to the overall magnitude of the 
extinction,
which we fix by demanding that a given cosmology reproduces the observed
luminosity distance at $z\sim 0.5$, i.e., that dust extinction can 
explain the dimming of the presently observed supernova sample, 
otherwise attributed to the ``concordance'' cosmology
$\Omega_{\rm M}=0.3$, $\Omega_\Lambda=0.7$.

The Monte-Carlo simulation program SNOC (\cite{SNOC})
was used to perform the integral in Eq.~(\ref{eq:deltam})
numerically  by following individual light paths through a large
number of cells containing galaxies and intergalactic dust. Through
each cell the background cosmology, the wavelength of the photon and the
dust density were updated, and the contributions from each cell added. Note that the model is 
approximately valid also for a patchy dust distribution, 
as long as the scale of 
inhomogeneities is small enough, i.e., $1/\sqrt{N}\ll 1$ where $N$ is the 
number of 
dust clouds intersected by the light-ray.

\begin{figure}
    \centerline{\includegraphics[width=\hsize]{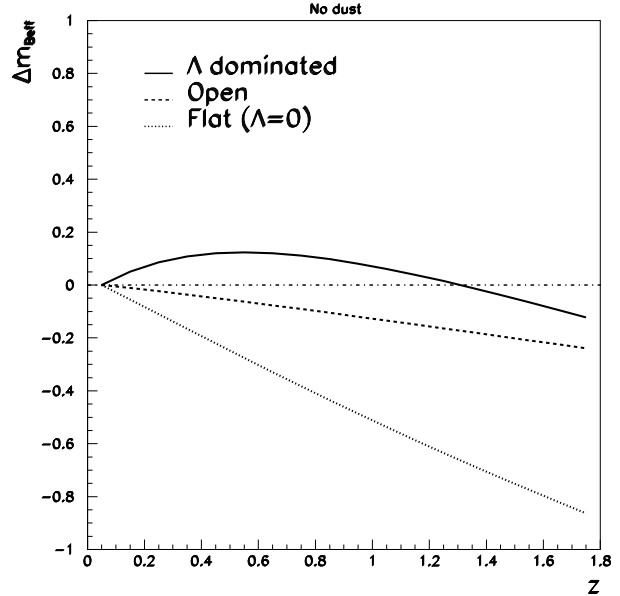}} 
    \caption[]{Differential magnitude for three cosmologies, 
$\Omega_{\rm M},\Omega_\Lambda$=(0.3,0.7) (solid line), (0.2,0) (dashed line)
and (1,0) (dotted line), compared 
with an empty universe, ($\Omega_{\rm M},\Omega_\Lambda$)=(0,0) 
(horizontal, dash-dotted line).}
    \label{fig:hubble1}
\end{figure}

First we summarize the situation concerning the dependence of
the observed magnitudes on the cosmological background model, in
the absence of dust. Figure \ref{fig:hubble1} shows the magnitude difference  
for three cosmologies,
($\Omega_{\rm M},\Omega_\Lambda$)=(0.3,0.7), (0.2,0) and (1,0), compared 
with an empty (Milne) universe, ($\Omega_{\rm M},\Omega_\Lambda$)=(0,0). 
In a matter dominated universe objects are increasingly 
brighter as a function of redshift in comparison with the
``empty'' universe. In a $\Lambda$-dominated universe, on the 
other hand, there is a turning point where the mass density 
overtakes the effect of the vacuum energy density above 
$z>1$.

Turning now to extinction due to a homogeneous component of dust,
we consider in this work the following scenarios: 

\begin{itemize}
\item Three combinations of cosmological parameters: 
($\Omega_{\rm M},\Omega_\Lambda$) = $(0.3,0.7),(0.2,0)$ and $(1,0)$.

\item $R_V$ ranging from 5.5 to 9.5, constant in the interval $0<z<2$. 

\item $\rho_{\rm dust} = \rho^\circ_{\rm dust}(1+z)^\alpha$, where 
\begin{eqnarray*}
 \alpha(z) =  
      \left\{ 
        \begin{array}{lc}
          3 {\rm \ for \ all \ z,}&  {\rm \ Model \ A} \\
          0 {\rm \ for \ z}>0.5 {\rm \ (3 \ for \ lower \  z).} &  {\rm \ Model \ B} 
       \end{array}
        \right.   
\end{eqnarray*}
Thus, Model B implies that $$\rho_{\rm dust}(z>0.5) = \rho_{\rm
dust}(z=0.5)= {\rm \ constant},$$ thus the comoving density increases
with cosmic time until $z=0.5$ from which it is constant.

%\left{= 3 {\rm model A} \\
%                       = 3 for z<0.5 and 0 for z>0.5 {\rm model B} \right.$
% 
% is either 3 for all redshifts, i.e. constant comoving
%   dust density (model A), 
%   or $\alpha(z \le 0.5)=3$ and $\alpha(z \ge 0.5)=0$ (model B).
 
\item The extinction scale-length, i.e., the interaction length for photon scattering
with dust particles, $\lambda \propto (\sigma \cdot \rho^\circ_{\rm dust})^{-1}$ is in the 
range (5 - 300)$\cdot\left({h \over 0.65}\right)$ Gpc, where $\sigma$ is the interaction cross-section.

\end{itemize}

Although  models A and B are not necessarily very plausible, they serve as useful limiting cases for 
more natural scenarios where the dust density is generated and distributed over a finite 
time scale.

A more detailed
calculation  would make use of an explicit model for the past star formation
rate. However, the predicted evolution of the comoving density between, say, 
$z\sim 0.8$ and $z\sim 1.7$  corresponds
to less than around a factor of 2 (\cite{madau}). 
%As an extreme example of cosmic evolution
%of the comoving dust density we will also consider the case of constant
%physical density, $\rho_{dust} =\rho_0={\rm const.}$.

\begin{figure}
  \begin{center}
    \includegraphics[width=\hsize]{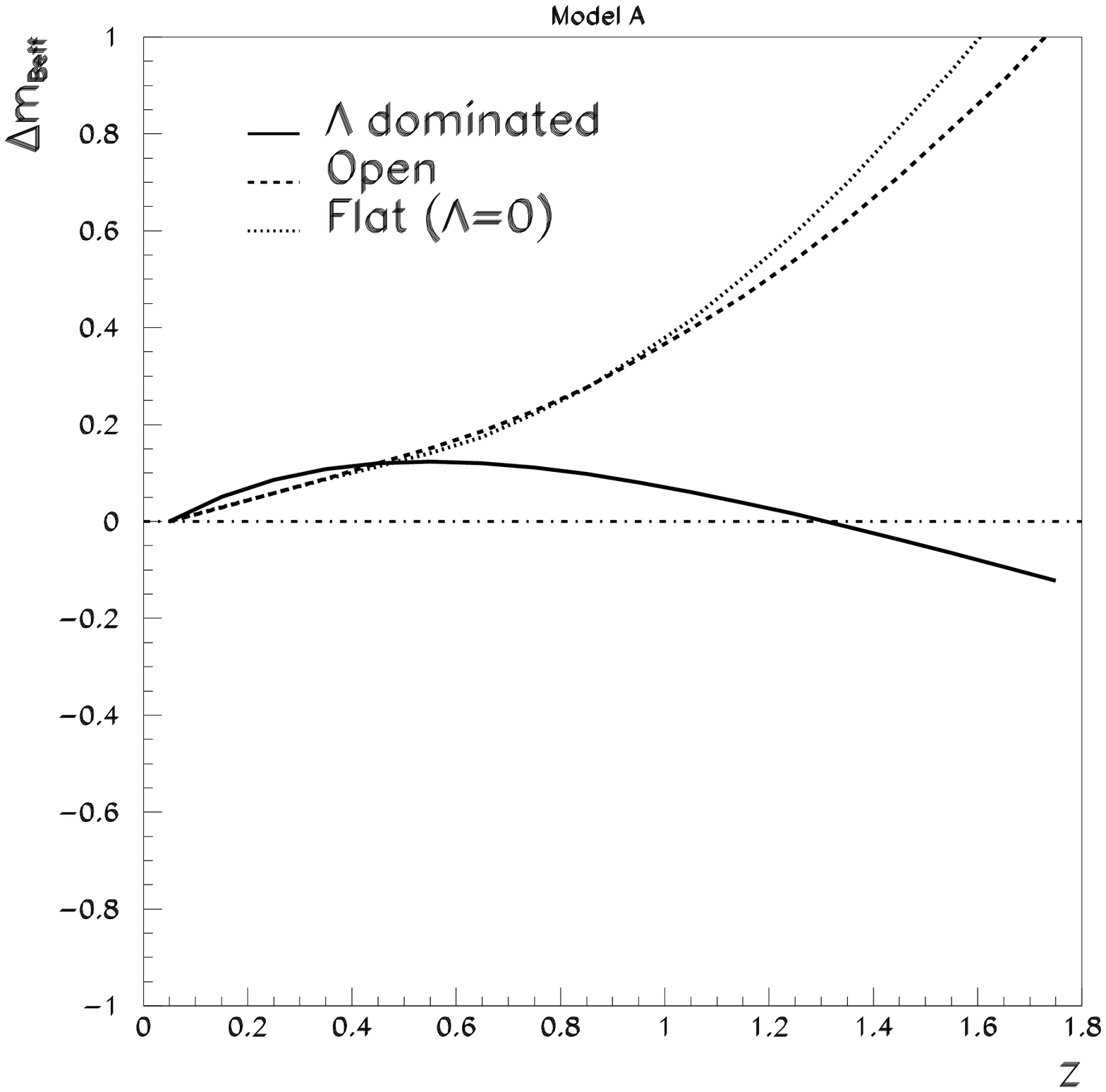} \\
     \includegraphics[width=\hsize]{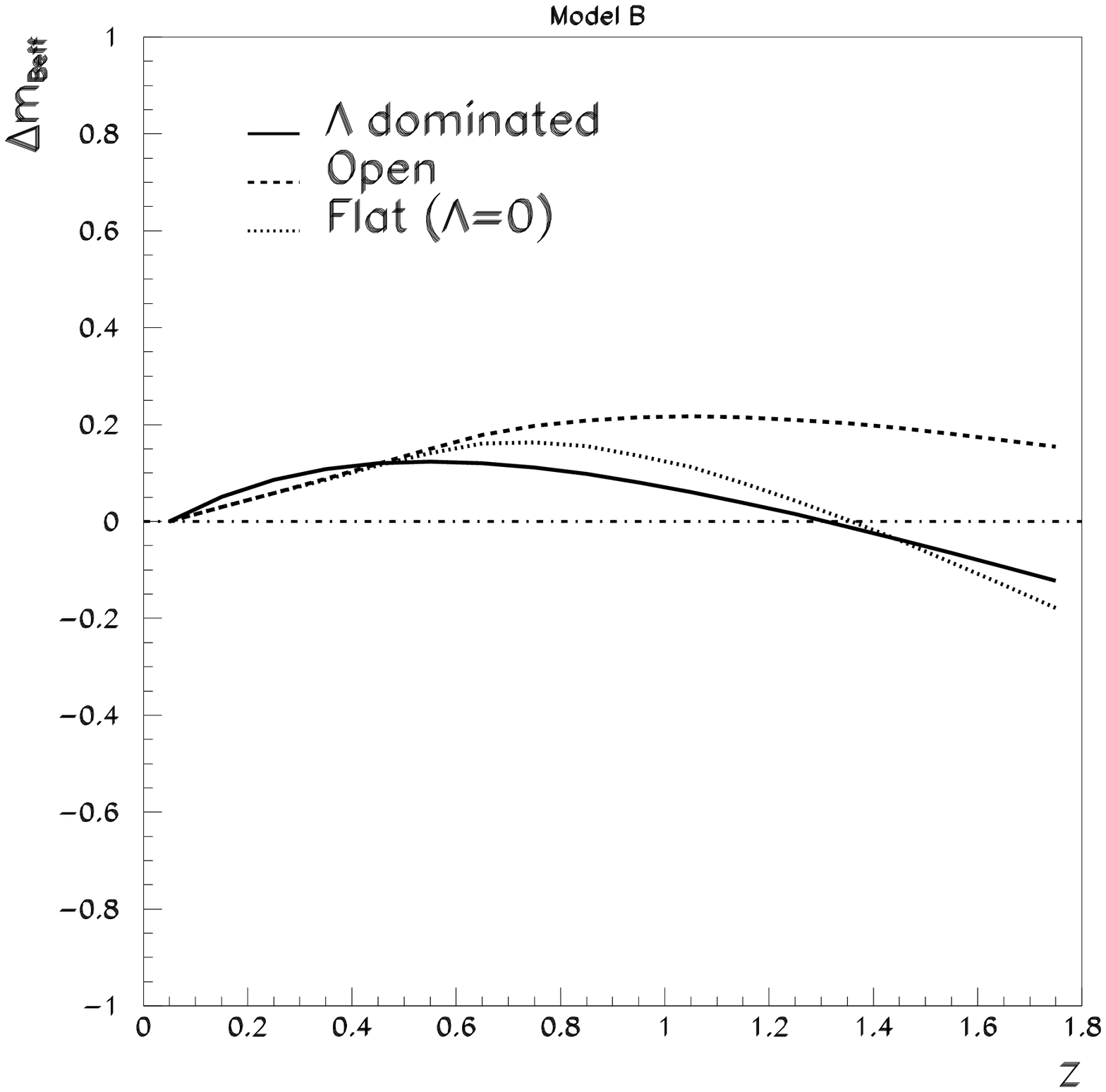} \\
     \caption{Differential magnitude for three cosmologies,
     $\Omega_{\rm M},\Omega_\Lambda$=(0.3,0.7), (0.2,0) and (1,0), compared
     with an empty universe, $\Omega_{\rm M},\Omega_\Lambda$=(0,0), where
     the faintness of the supernovae at $z\sim 0.5$ is achieved by
     introducing a homogeneous density of intergalactic dust with
     $R_V$=5.5. Upper panel: dust model A. Lower panel: dust model B
     (see text for details).}  \label{fig:hubble2} \end{center}
\end{figure}

In Fig.\,~\ref{fig:hubble2}, the apparent
faintness of the high-redshift supernovae at $z\sim0.5$ is forced by 
introducing a homogeneous grey-dust component. Contrary to the conclusions in (\cite{riess17}), we find that 
the restframe B-band brightness of SN1997ff cannot be used to reliably
exclude the presence of extragalactic dust at high redshift. The
assumption in their analysis is that extinction can be
parametrized as $\Delta m_{\rm Beff} = 0.3 \cdot z$ mag in a 
$(\Omega_{\rm M},\Omega_\Lambda)=(0,0)$ universe. This falls 
in between our models A and B. If we consider
model B on the other hand, there is less than 0.1 mag difference at $z\sim 1.7$ between
the $\Lambda$-dominated and what would be observed in a $\Omega_{\rm M}=1$ universe 
with the required dust
density to explain the Type Ia supernova measurements at $z\sim 0.5$,
as shown in lower panel of Fig.~\ref{fig:hubble2}.  

Clearly, accurate
spectrophotometric data from a statistical sample of supernovae at
this redshift are needed to settle the issue, as described below.

%\begin{enumerate}
%\item constant comoving density, $\rho(z)=\rho_0 (1+z)^3$ (model A). 
%\item  $\rho(z)=\rho_0 (1+z)^3$ for z$\le 0.5$ and  $\rho(z>0.5) = \rho(0.5)$   
%\end{enumerate}

\subsection{Spectrophotometry in the presence of dust}

\begin{figure}
\centerline{\includegraphics[width=\hsize]{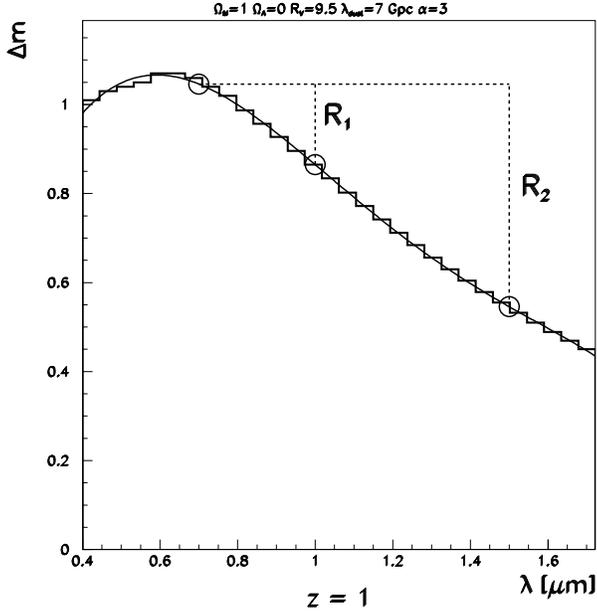}} 
  \caption[]{Monte-Carlo simulation of the differential extinction between 
$\lambda=0.4$ and $1.7$ $\mu$m for a source at a redshift $z=1$. The solid line shows a fit
to a fifth order polynomial. $R_1$ is defined to be the magnitude difference
in extinction between 0.7 $\mu$m and 1.0 $\mu$m.   
$R_2$ measures the differential extinction between 0.7 $\mu$m and 1.5 $\mu$m.}
  \label{fig:colordef} 
\end{figure}

In order to quantify the induced differential spectral shift due to large dust grains
we introduce two quantities: $R_1$ and $R_2$. 
$R_1$ is the magnitude difference
in extinction between $\lambda=0.7$ and $1.0$ $\mu$m.   
$R_2$ measures the differential extinction  between $\lambda=0.7$ and $1.5$ $\mu$m, 
as shown in Fig.\,~\ref{fig:colordef} for a source at $z=1$. Also shown in the figure is 
a fit of a fifth order polynomial over the wavelength range $\lambda=0.4$ to $1.7$ $\mu$m. 

The rationale for introducing $R_1$ and $R_2$ is twofold: 1) the color dependence
is almost linear in the range between $\lambda=0.7$ and $1.5$ $\mu$m. 2) $R_1$ and $R_2$
can be measured within the sensitive range of an infrared  (HgCdTe) detector. $R_1$ can 
{\em also} be measured with a CCD detector.

\section{Grey dust simulations}
\subsection{Open or flat universe with vanishing $\Lambda$}

The idea of Aguirre (\cite{aguirre99a,aguirre99b}) was to explain the faintness of high-z Type Ia
supernovae without invoking a ``Dark Energy'' component. In such a scenario, an 
extinction of $0.2-0.5$ magnitudes is required to reconcile the supernova data with either
an open or flat universe without a cosmological constant. Although perfectly
grey dust (i.e., with wavelength-independent absorption) with a fine-tuned
redshift distribution possibly can mimick the effects of a cosmological
constant, we want to investigate here whether a more natural  dust 
model can do the same. The point to make is that realistic dust firstly
has to be related to astrophysical sources, such as star formation, and
secondly that it always implies {\em some} wavelength-dependence in the
absorption and scattering properties.

\begin{figure}
  \includegraphics[width=5.1cm]{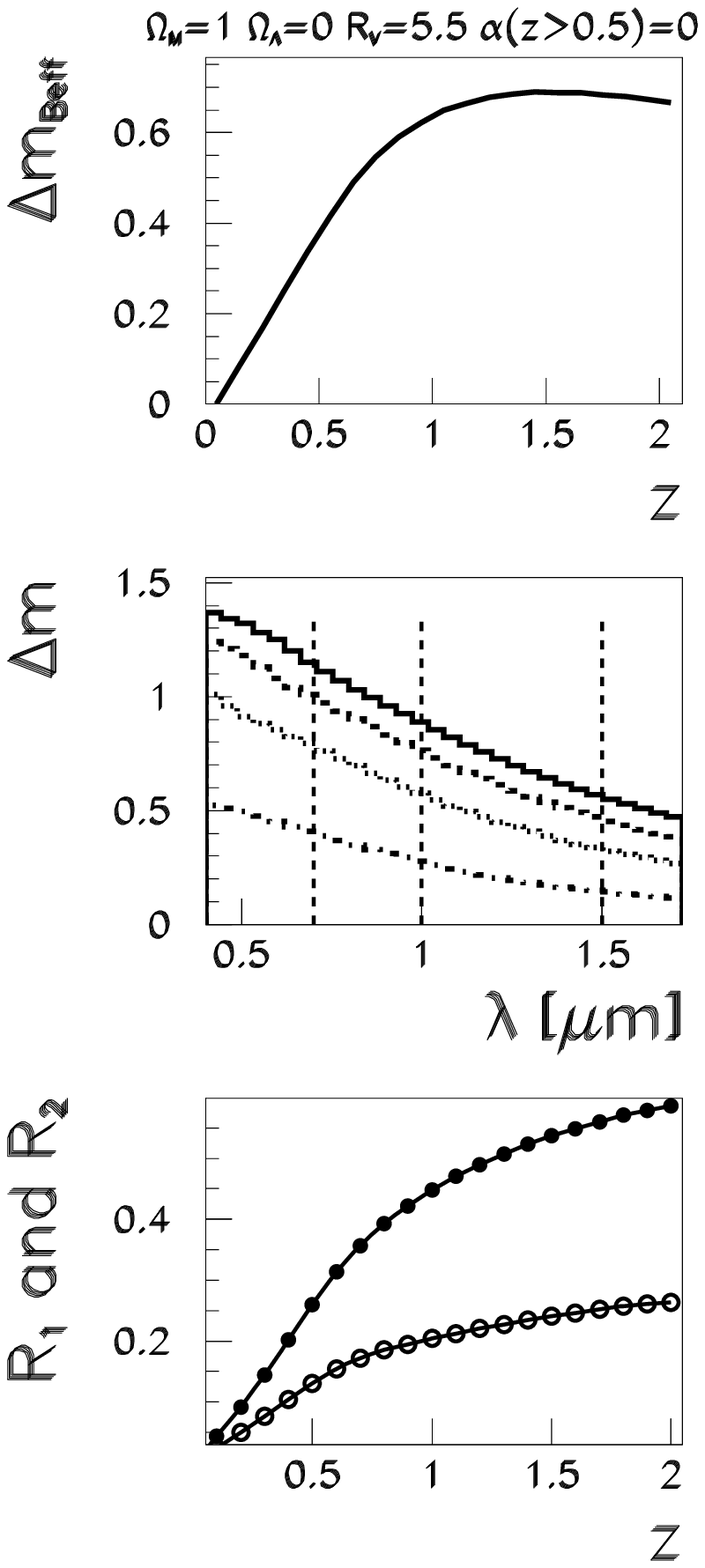} \hspace*{-1.6cm} 
  \includegraphics[width=5.1cm]{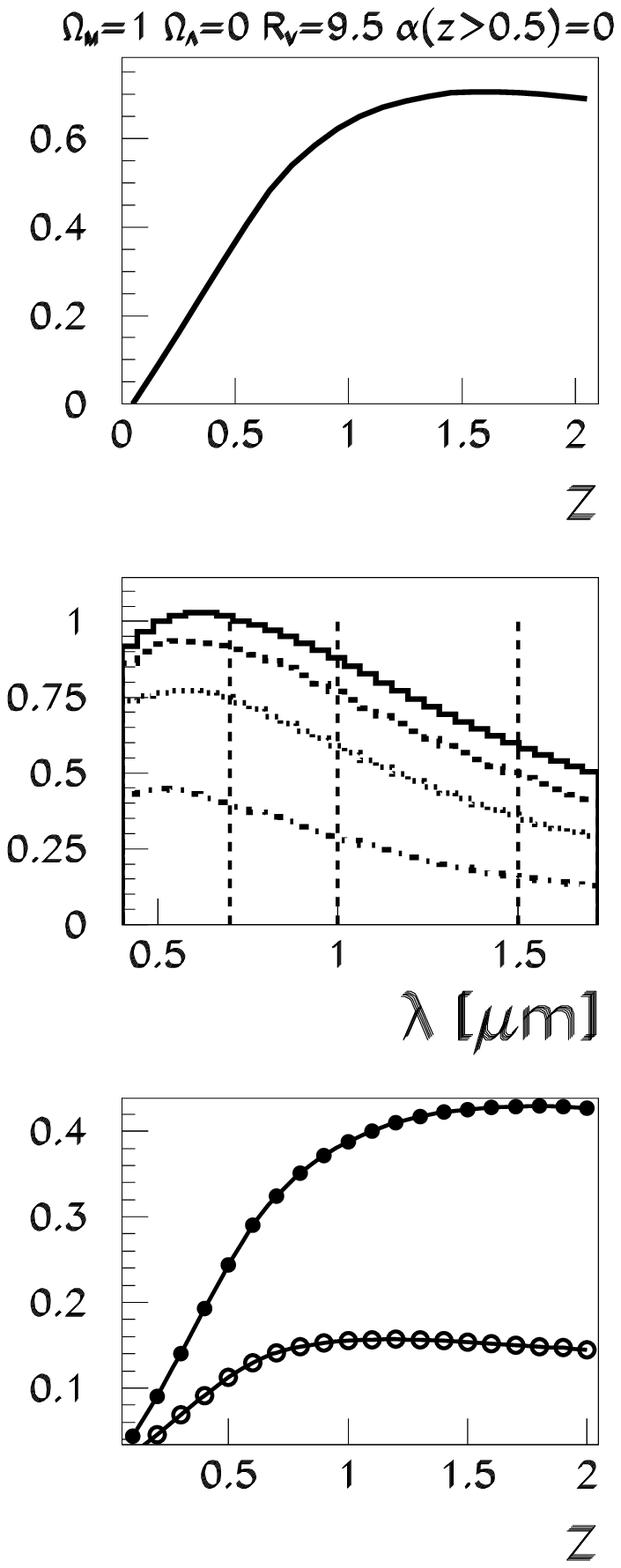} 
  \caption[]{Dust extinction in model B for $R_V$=5.5 (left side) and $R_V$=9.5 (right side) 
in a flat ($\Omega_{\rm M}$=1) universe with the dust density 
adjusted to generate the required dimming of the restframe B-band magnitude from a source 
at at $z=0.5$. The upper panels show the dimming in the observational band corresponding to 
the restframe B-band. The middle panels show the differential extinction as a function of wavelength
for a source at redshifts $z=0.5,1.0,1.5$ and $2.0$. The dashed lines show the position of $\lambda=
0.7, 1.0$ and 1.5 $\mu$m used to calculate $R_1$ and $R_2$. The bottom panels show the differential
color coefficients $R_1$ and $R_2$ as a function of redshift.}
  \label{fig:om1p0oxp0rv5p5_2} 
\end{figure}

\begin{figure}
  \includegraphics[width=5.1cm]{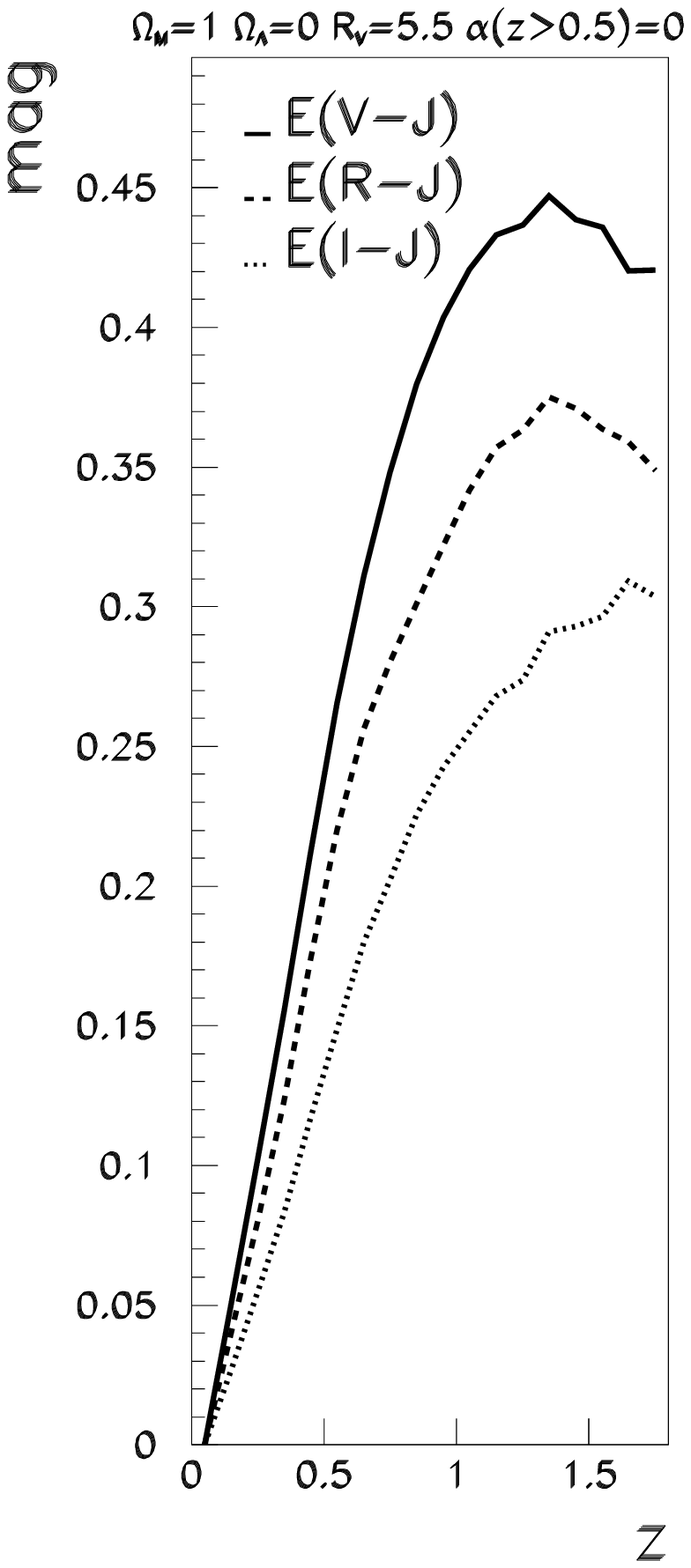} \hspace*{-1.6cm} 
  \includegraphics[width=5.1cm]{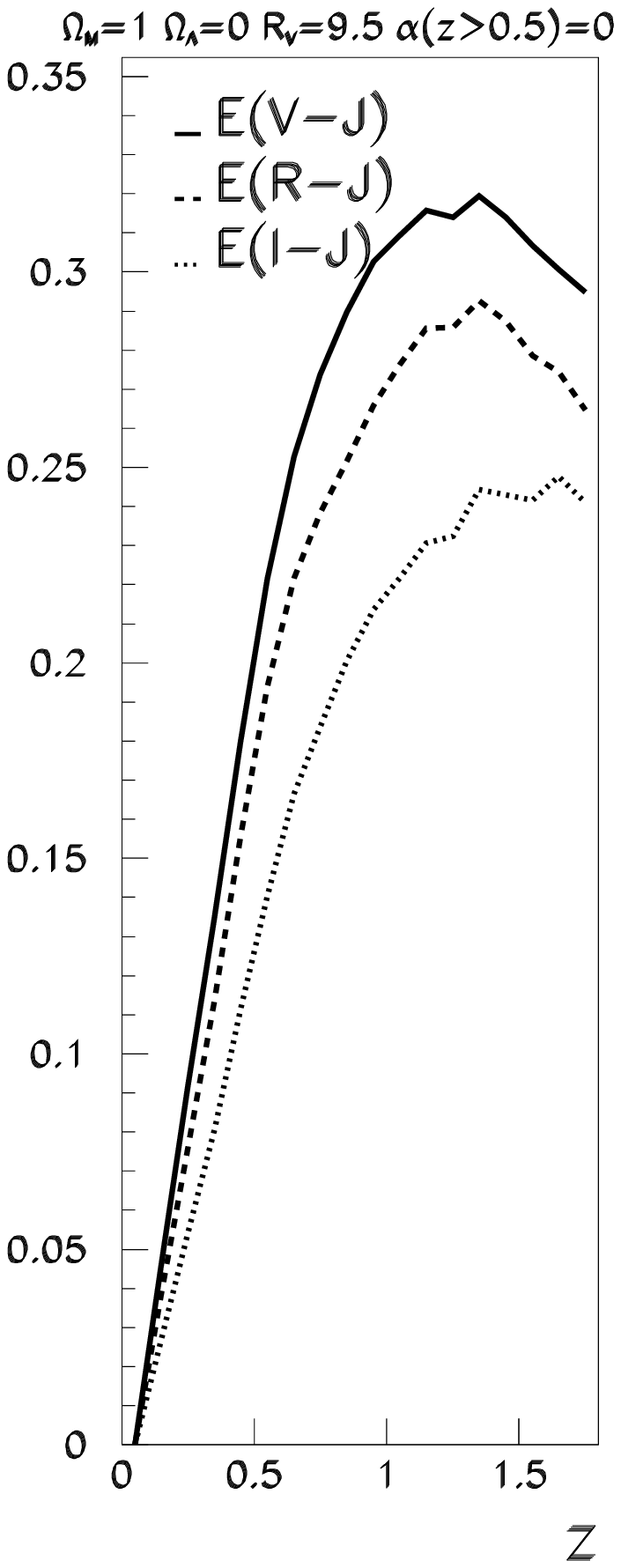} 
\caption[]{Color extinctions E(V-J),E(R-J) and E(I-J) for Type Ia supernovae in dust model B
 for $R_V=5.5$ (left side) and $R_V=9.5$ (right side) 
in a flat ($\Omega_{\rm M}$=1) universe with the dust density 
adjusted to generate the required dimming of the restframe B-band magnitude from a source 
at at $z=0.5$.}
\label{fig:color_om1p0oxp0rv5p5_2}
\end{figure}

Fig.\,~\ref{fig:om1p0oxp0rv5p5_2} shows the magnitude and differential
extinction as a function of source redshift for model B and $R_V=5.5$
and $9.5$ in a flat universe without cosmological constant. The color
terms for the highest redshifts are $R_1 \approx 0.2$ and $R_2 \approx
0.6$, i.e., clearly measurable with a population of SNe as expected
for the proposed SNAP
satellite. Fig.\,~\ref{fig:color_om1p0oxp0rv5p5_2} shows the color
extinction (at maximum intensity) in V-J, R-J and I-J for a normal
Type Ia supernova as a function of redshift\footnote{All the broadband
filters and spectroscopy walvelength scales are in the observer's
frame}.  At $z\sim 0.5$, where several supernovae have been observed
in the NIR, e.g. (\cite{riessNIR}), approximately $0.1-0.15$ and
$0.2-0.25$ magnitude extinction is to be expected in I-J and V-J
respectively.

\begin{figure}
  \includegraphics[width=5.1cm]{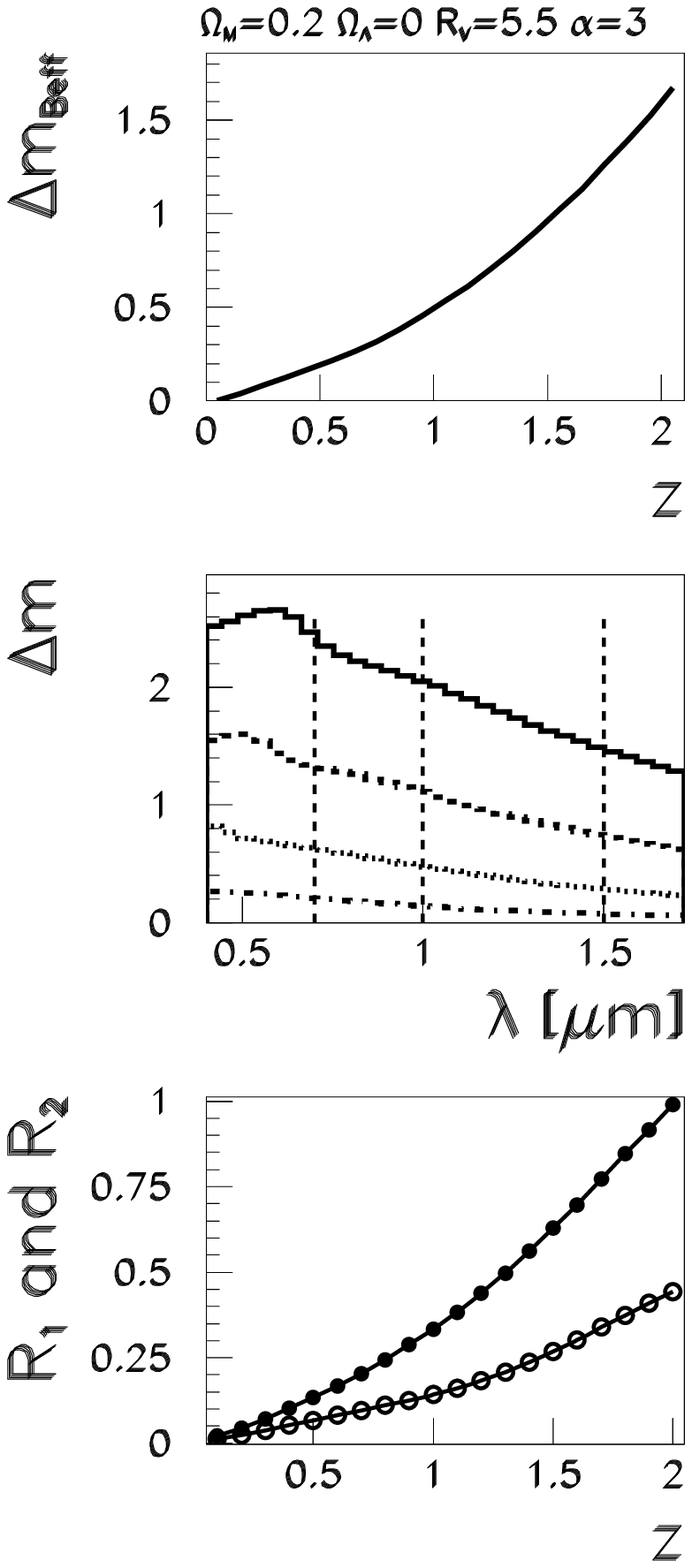} \hspace*{-1.6cm} 
  \includegraphics[width=5.1cm]{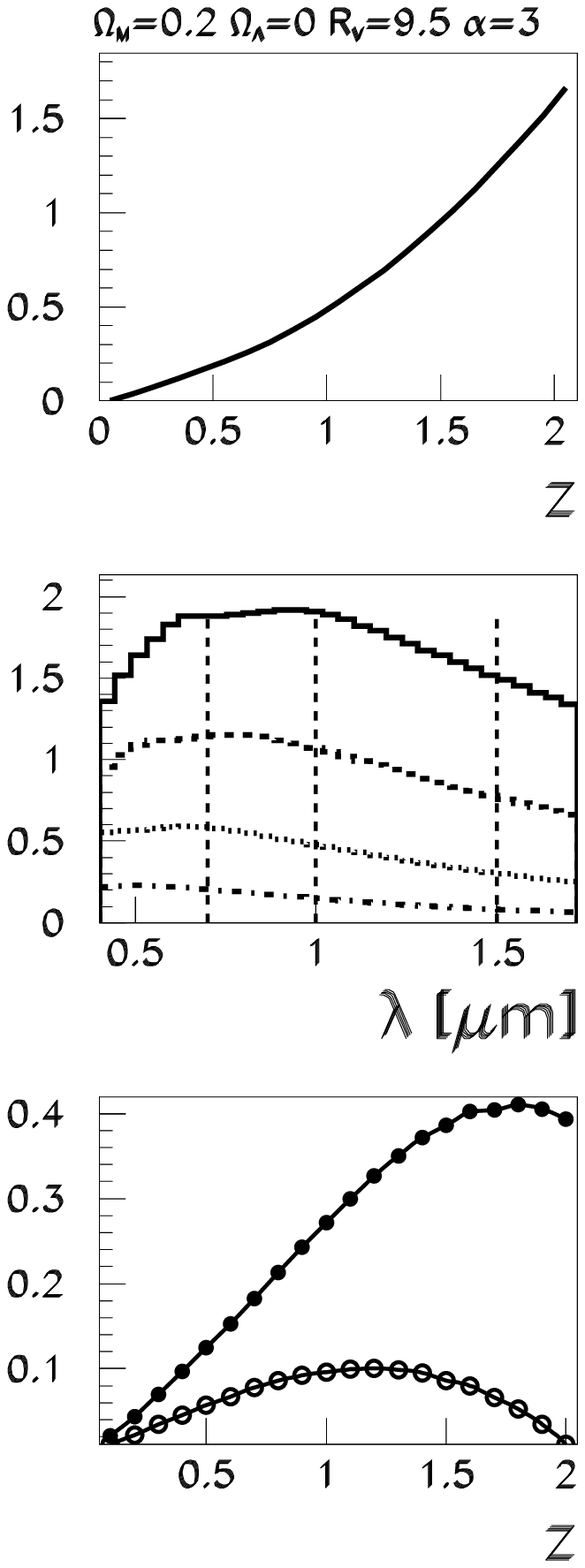} 
  \caption[]{Dust extinction in model A for $R_V=5.5$ (left side) and $R_V=9.5$ (right side) 
in an open ($\Omega_{\rm M}=0.2$) universe with the dust density 
adjusted to generate the required dimming of the restframe B-band magnitude from a source 
at at $z=0.5$. The upper panels show the dimming in the observational band corresponding to 
the restframe B-band. The middle panels show the differential extinction as a function of wavelength
for a source at redshifts $z=0.5,1.0,1.5$ and $2.0$. The dashed lines show the position of $\lambda=
0.7, 1.0$ and 1.5 $\mu$m used to calculate $R_1$ and $R_2$. The bottom panels show the differential
color coefficients $R_1$ and $R_2$ as a function of redshift.}
  \label{fig:omp2oxp0rv5p5_1} 
\end{figure}

\begin{figure}
 \includegraphics[width=5.1cm]{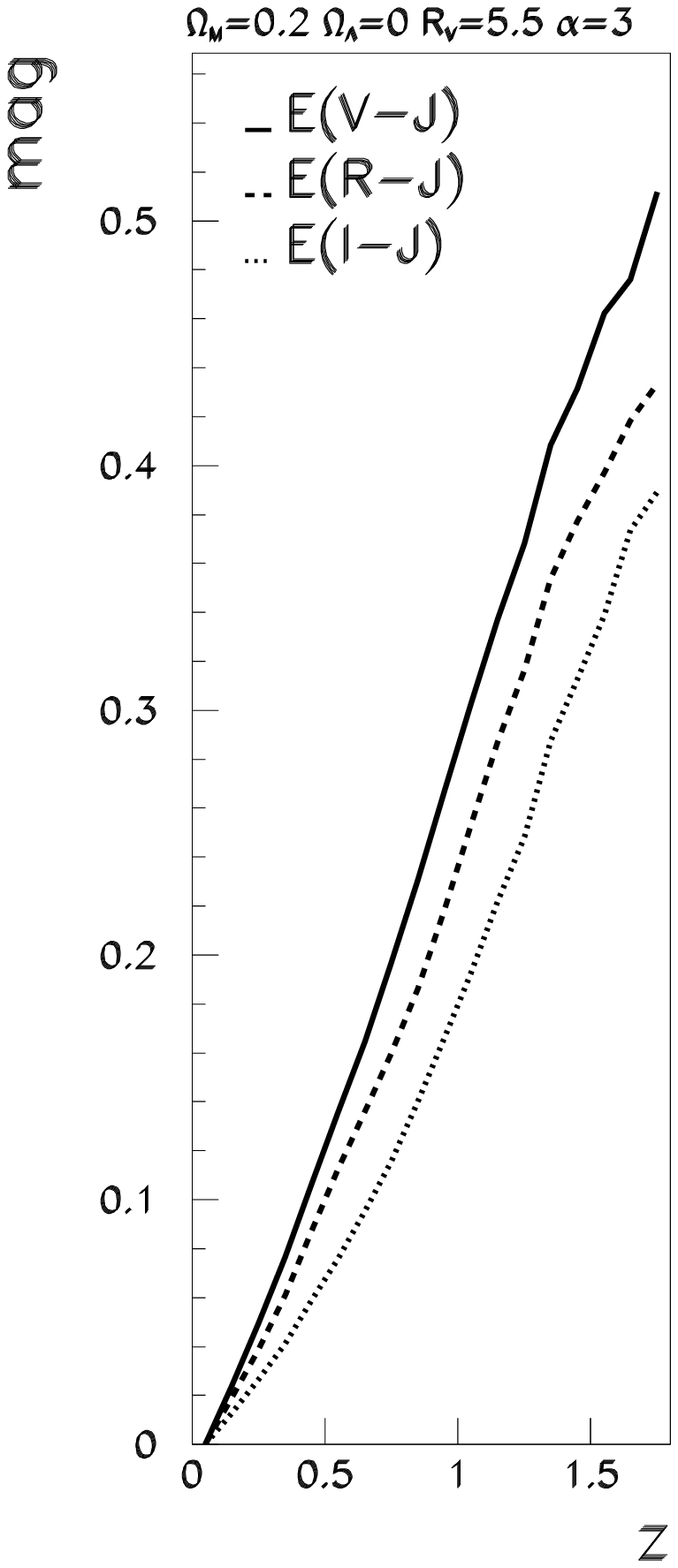} \hspace*{-1.6cm} 
 \includegraphics[width=5.1cm]{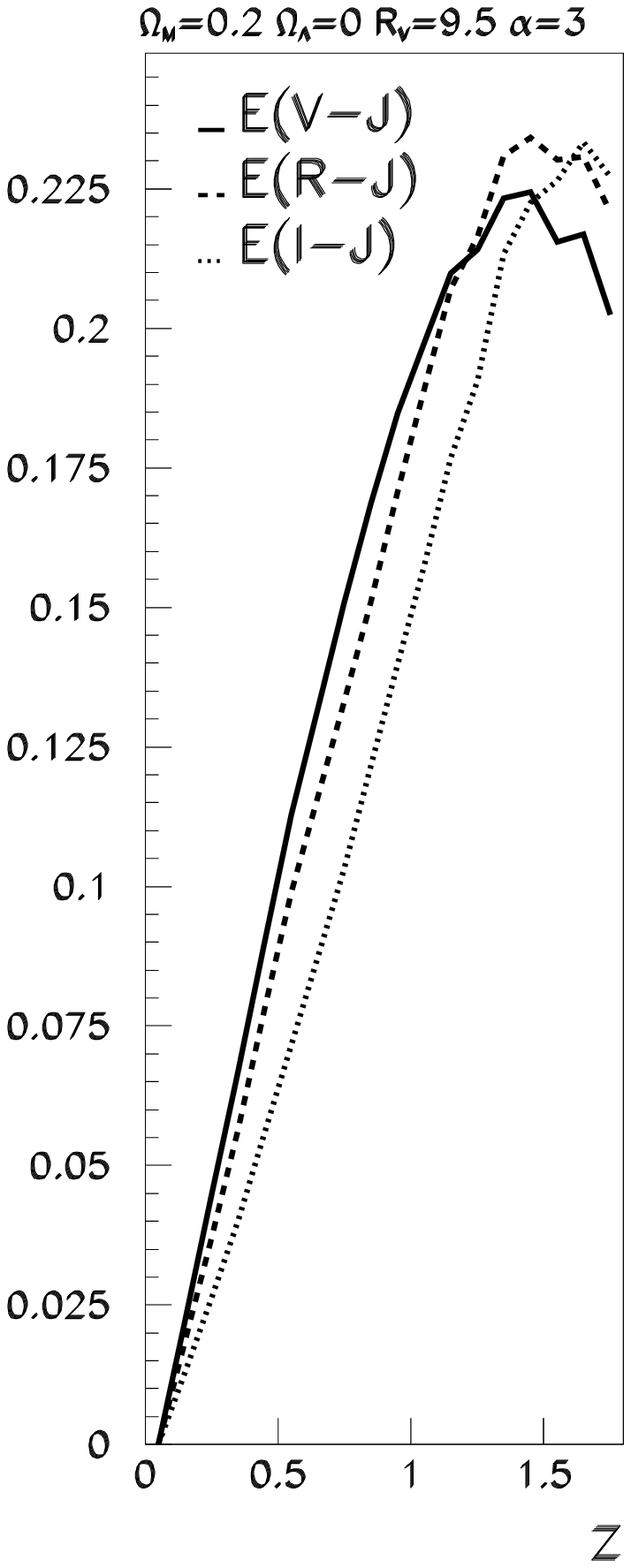}
\caption[]{Color extinctions E(V-J),E(R-J) and E(I-J) for Type Ia supernovae in dust model A
 for $R_V=5.5$ (left side) and $R_V=9.5$ (right side) 
in an open ($\Omega_{\rm M}=0.2$) universe with the dust density 
adjusted to generate the required dimming of the restframe B-band magnitude from a source 
at at $z=0.5$.}
\label{fig:color_omp2oxp0rv5p5_1}
\end{figure}

Fig.\,~\ref{fig:omp2oxp0rv5p5_1} shows the effects of dust extinction in an open universe
for $R_V=5.5$ and 9.5 where the dust density has been adjusted to fit the 
observational constraints at $z\sim 0.5$. In both figures, a constant comoving dust density
was used in the calculations, i.e. $\alpha=3$ (model A). Fig.\,~\ref{fig:color_omp2oxp0rv5p5_1}
shows the color extinction for normal Type Ia SNe for the same conditions as in 
Fig.\,~\ref{fig:omp2oxp0rv5p5_1}.

\begin{figure}
\includegraphics[width=5.1cm]{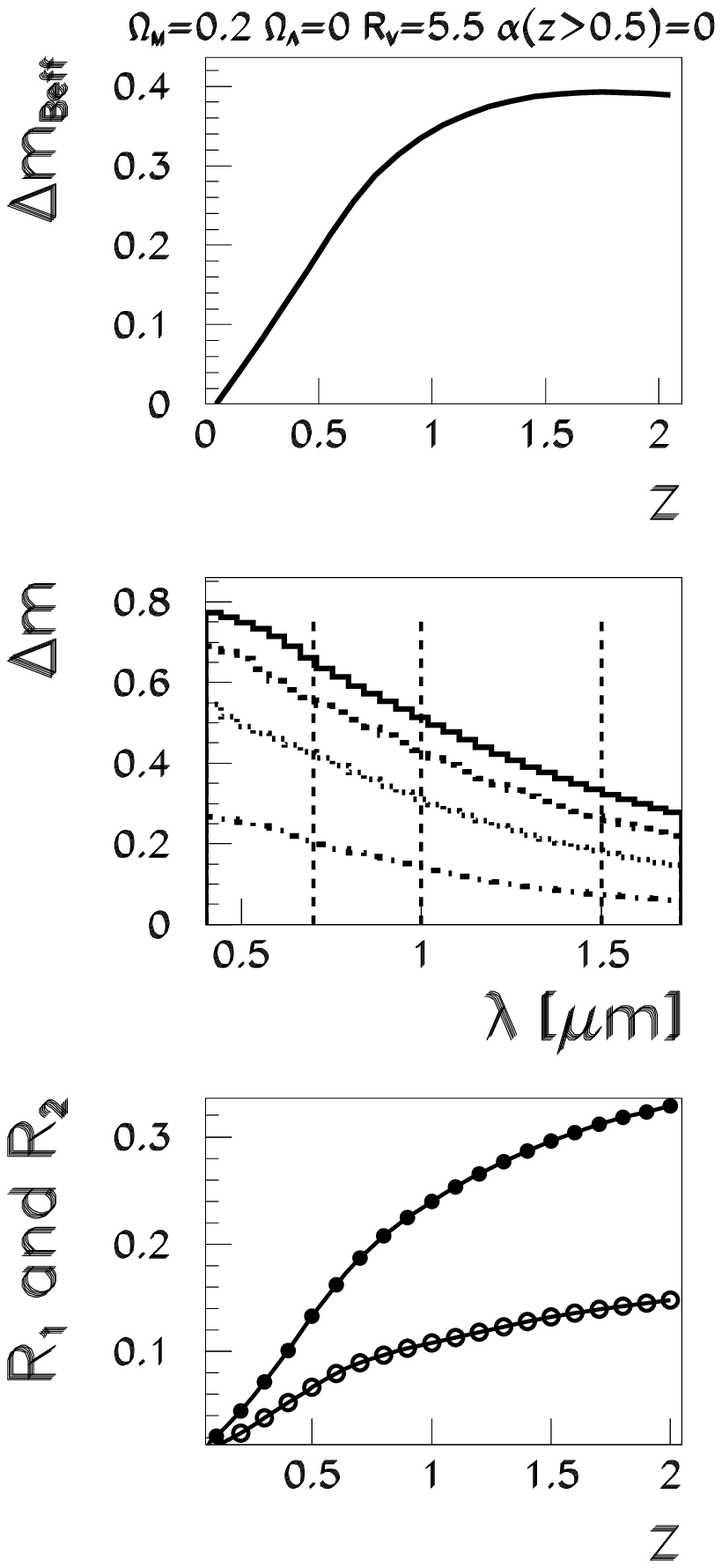}  \hspace*{-1.6cm} 
\includegraphics[width=5.1cm]{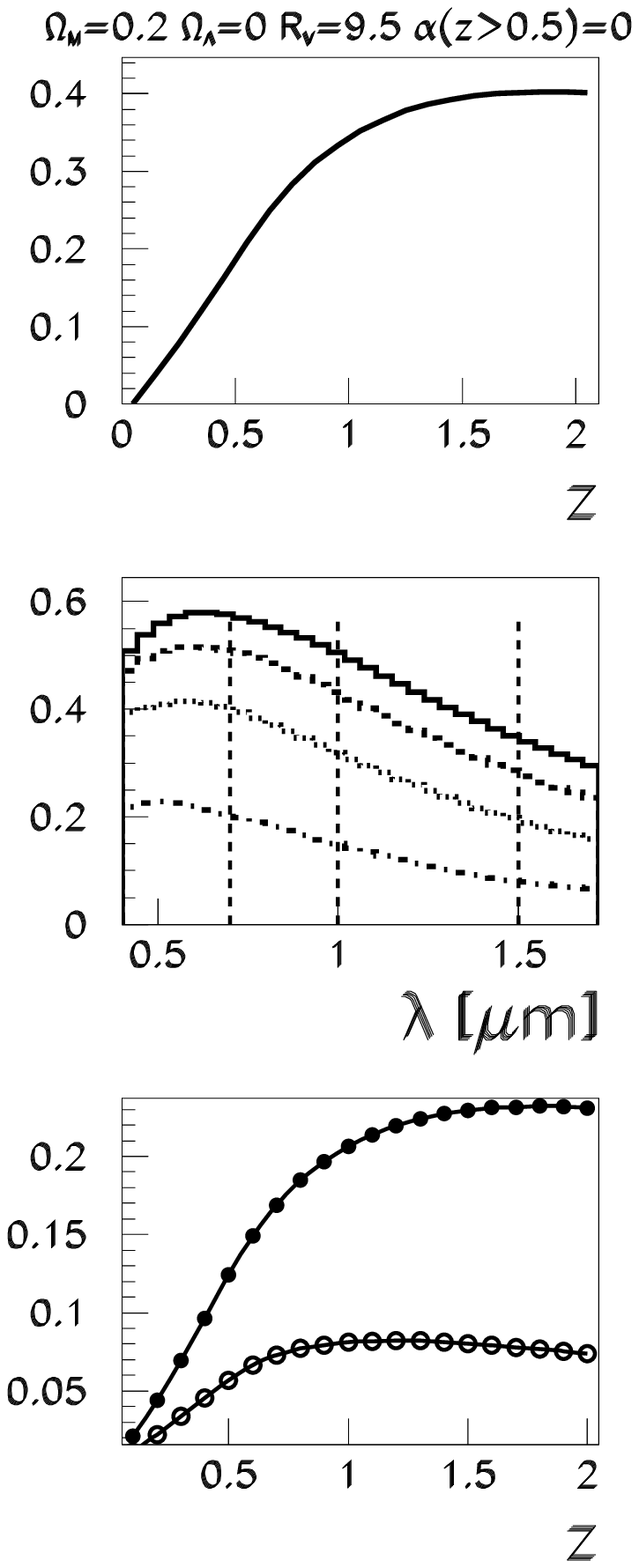} 
  \caption[]{Dust extinction in model B for $R_V$=5.5 (left side) and $R_V$=9.5 (right side) 
in an open ($\Omega_{\rm M}$=0.2) universe with the dust density 
adjusted to generate the required dimming of the restframe B-band magnitude from a source 
at at $z=0.5$. The upper panels show the dimming in the observational band corresponding to 
the restframe B-band. The middle panels show the differential extinction as a function of wavelength
for a source at redshifts $z=0.5,1.0,1.5$ and 2.0. The dashed lines show the position of $\lambda$=
0.7, 1.0 and 1.5 $\mu$m used to calculate $R_1$ and $R_2$. The bottom panels show the differential
color coefficients $R_1$ and $R_2$ as a function of redshift.}
  \label{fig:omp2oxp0rv9p5_2} 
\end{figure}

\begin{figure}
  \includegraphics[width=5.1cm]{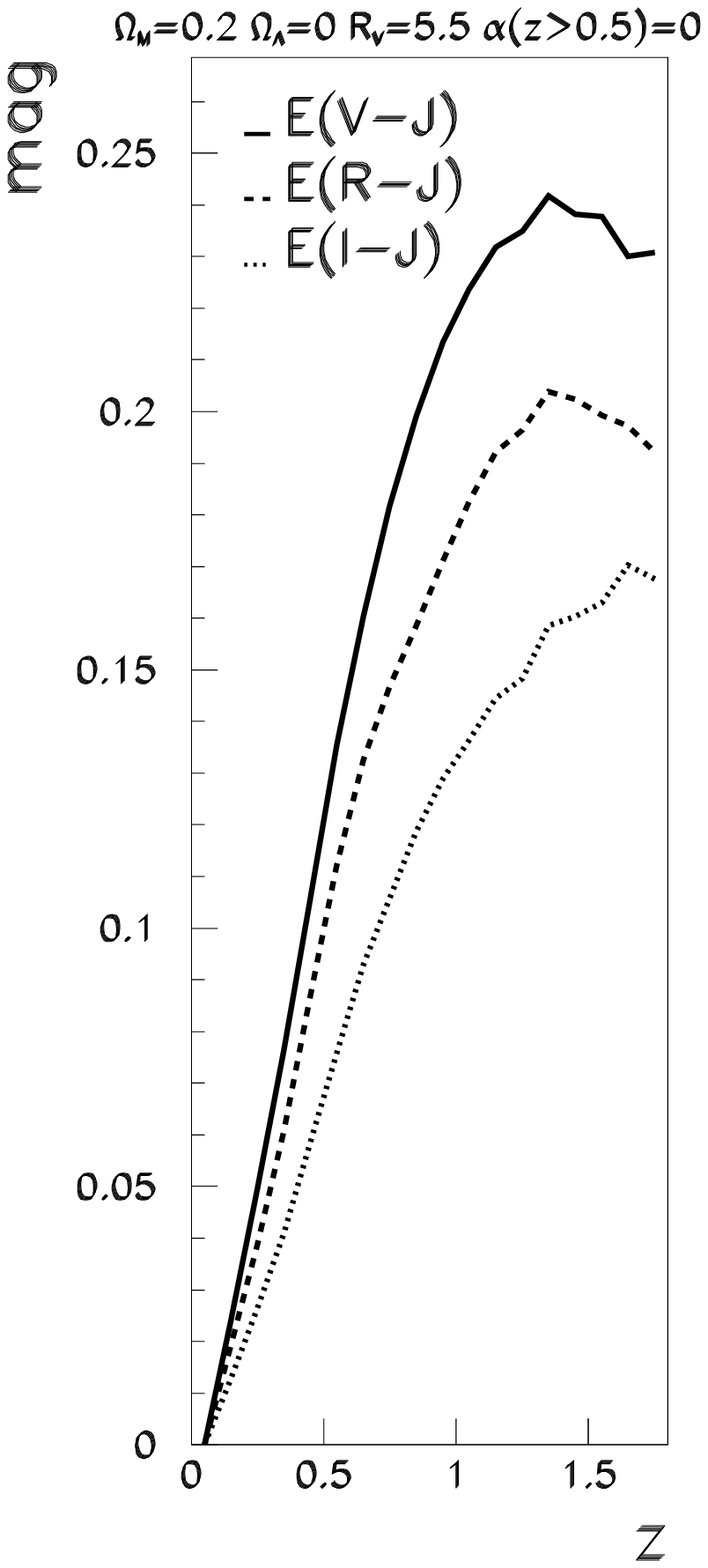} \hspace*{-1.6cm} 
  \includegraphics[width=5.1cm]{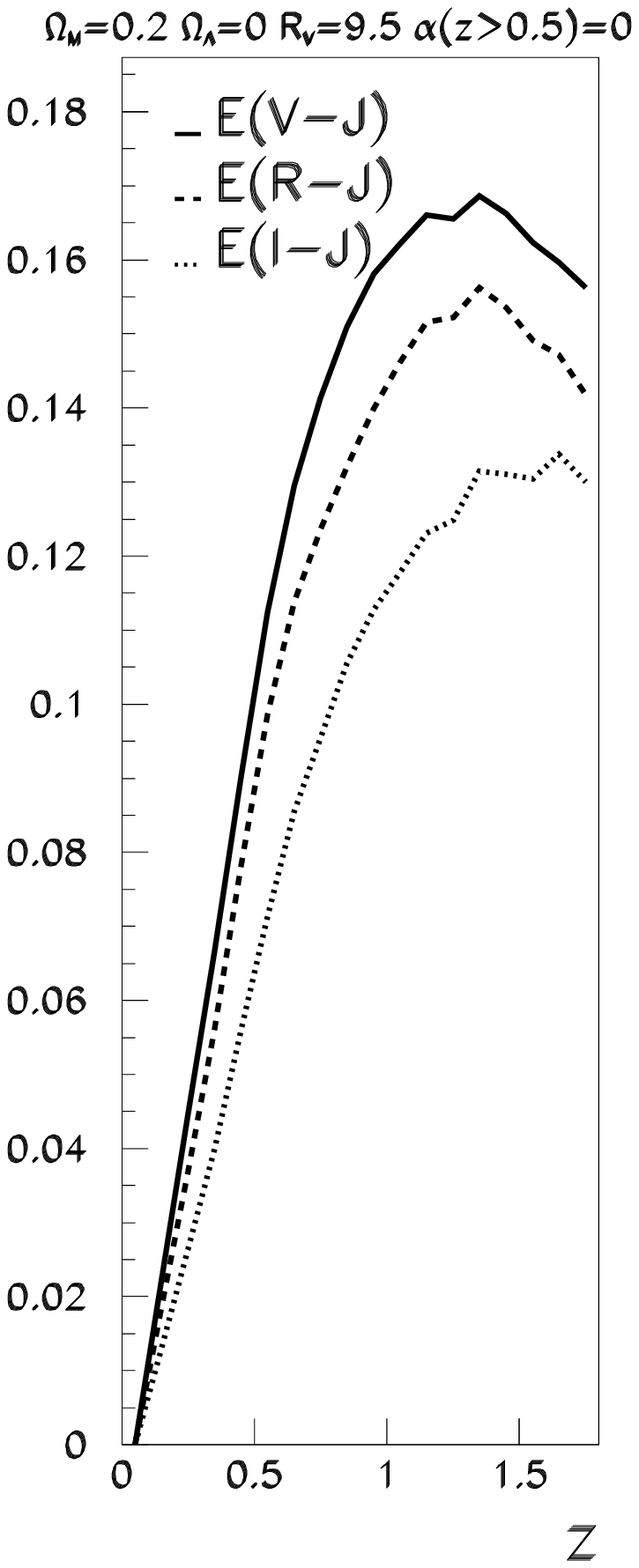}
\caption[]{Color extinctions E(V-J),E(R-J) and E(I-J)  for Type Ia supernovae in dust model B 
 for $R_V$=5.5 (left side) and $R_V$=9.5 (right side) 
in an open ($\Omega_{\rm M}$=0.2) universe with the dust density 
adjusted to generate the required dimming of the restframe B-band magnitude from a source 
at at $z=0.5$.}
\label{fig:color_omp2oxp0rv9p5_2}
\end{figure}

Figs.\,~\ref{fig:omp2oxp0rv9p5_2} and \ref{fig:color_omp2oxp0rv9p5_2} show the  effect of 
extinction in an open universe with dust model B.
Note that as all the models have been adjusted to match the brightness of 
Type Ia supernovae at $z=0.5$, 
there is no loss of generality in choosing the redshift at which
the dust density reaches its maximum at $z_{\rm max} = 0.5$ in model B. If $z_{\rm max}$ is increased,
the results will become more similar to model A. If $z_{\rm max}$ is decreased, the
differential extinction decreases somewhat. For example, for $z_{\rm max}=0.2$ in an open
universe, the 
shapes of the the curves in  Figs.\,~\ref{fig:omp2oxp0rv9p5_2} and \ref{fig:color_omp2oxp0rv9p5_2}
are essentially preserved while the absorption magnitude is approximately half as large 
as for  $z_{\rm max}=0.5$. If the redshift evolution of the dust density is more rapid, i.e. 
if there is a burst of production of intergalactic dust, it would be reflected as a sharp
discontinuity in the Hubble diagram of Type Ia supernovae and their colors.

%\subsubsection{Flat universe without cosmological constant}

\subsection{The ``standard'' $\Lambda$ dominated universe}

\begin{figure}
  \includegraphics[width=5.1cm]{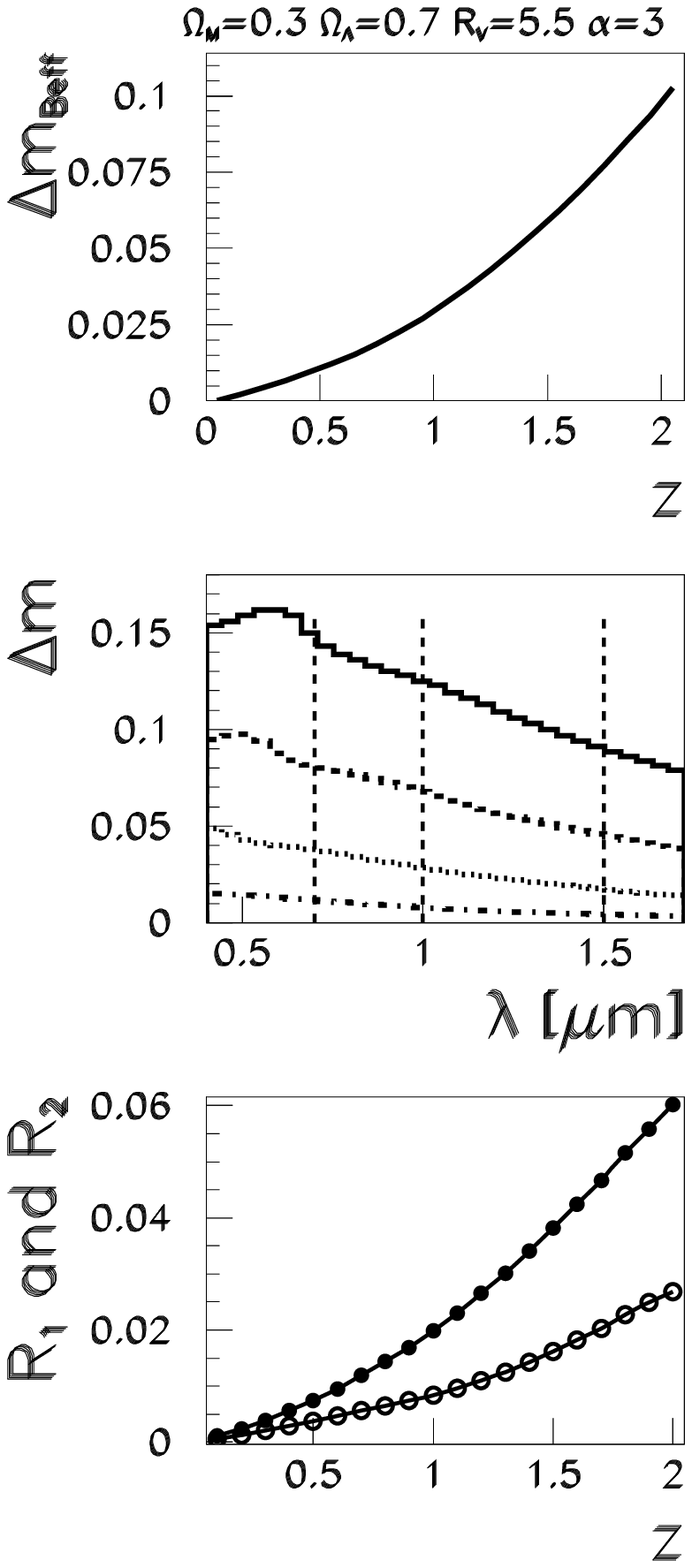} \hspace*{-1.6cm} 
  \includegraphics[width=5.1cm]{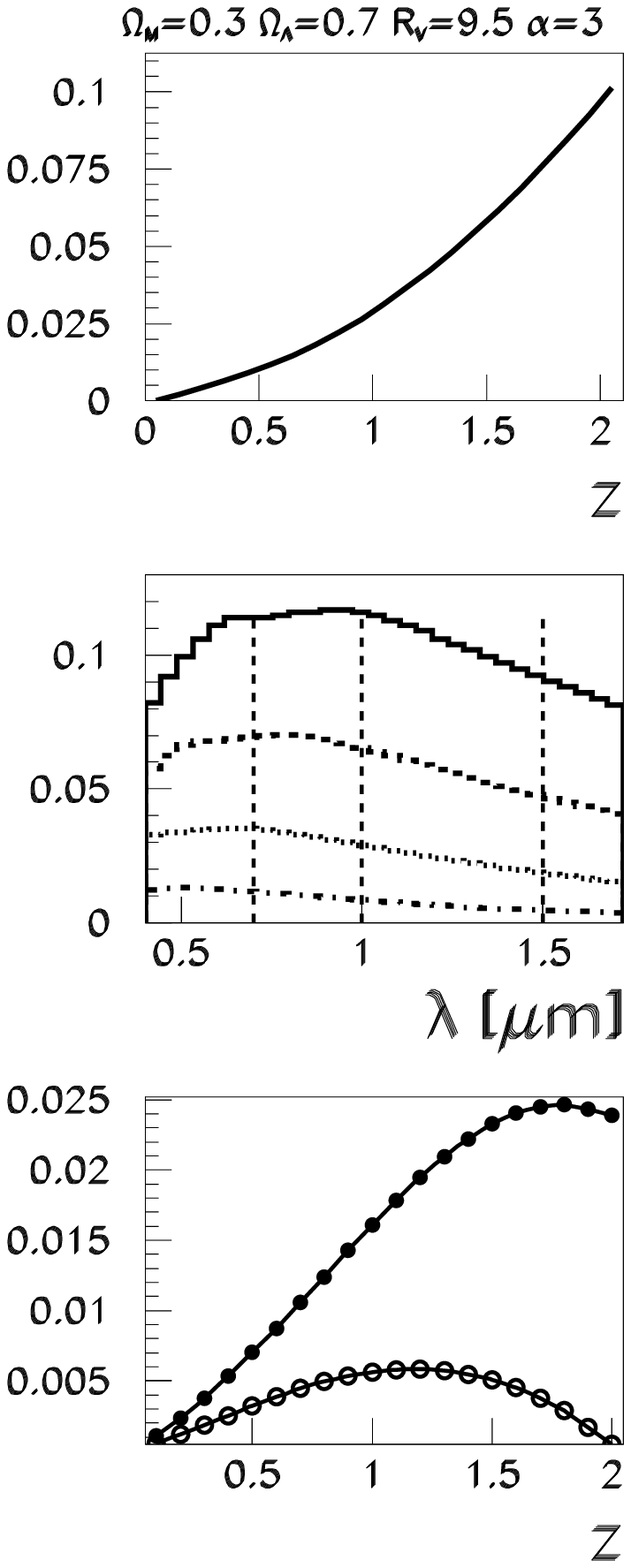} 
 \caption[]{Dust extinction in model A for $R_V$=5.5 (left side) and $R_V$=9.5 (right side) 
in a flat $\Lambda$-dominated universe with the dust density 
adjusted to generate about 0.1 mag dimming in the observed band corresponding to the 
restframe B-band magnitude from a source the the limiting SNAP depth 
($\lambda_{\rm dust}$=300$\cdot\left({0.65 \over h}\right)$ Gpc). 
The upper panels show the dimming in the observational band corresponding to 
the restframe B-band. The middle panels show the differential extinction as a function of wavelength
for a source at redshifts $z=0.5,1.0,1.5$ and 2.0. The dashed lines show the position of $\lambda$=
0.7, 1.0 and 1.5 $\mu$m used to calculate $R_1$ and $R_2$. The bottom panels show the differential
color coefficients $R_1$ and $R_2$ as a function of redshift.}
  \label{fig:omp3oxp7rv5p5_1} 
\end{figure}

\begin{figure}
  \includegraphics[width=5.1cm]{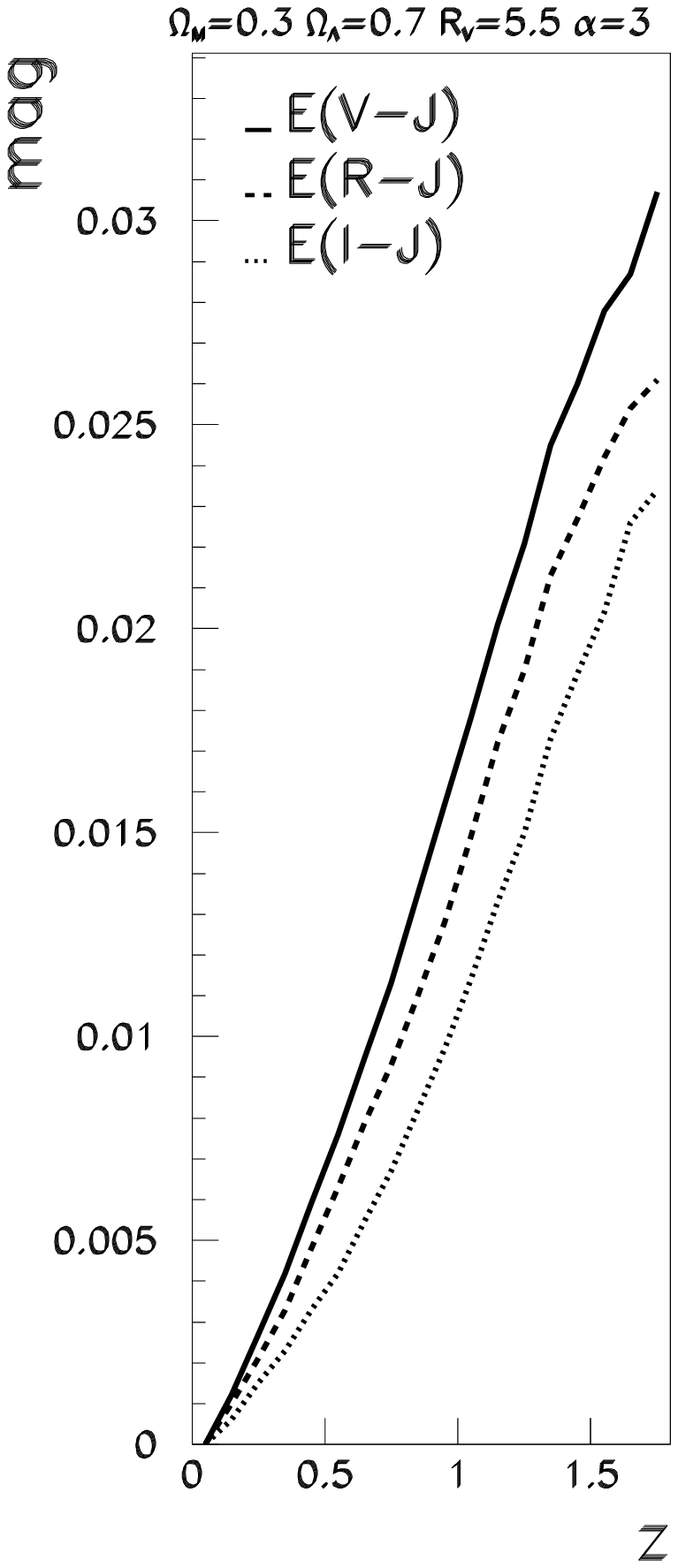} \hspace*{-1.6cm} 
  \includegraphics[width=5.1cm]{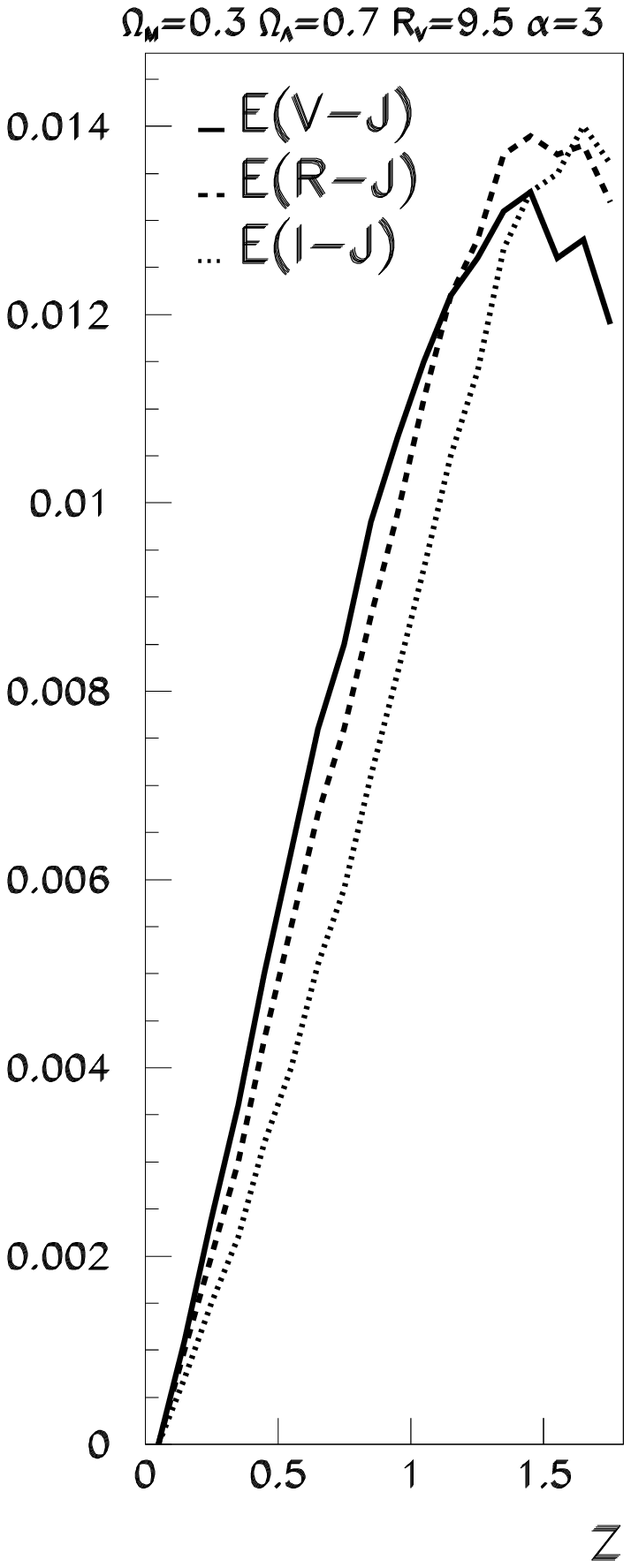}  
\caption[]{Color extinctions E(V-J),E(R-J) and E(I-J)  for Type Ia supernovae in dust model A 
 for $R_V$=5.5 (left side) and $R_V$=9.5 (right side) 
in a flat $\Lambda$-dominated universe with the dust density 
adjusted to generate about 0.1 mag dimming in the observed band corresponding to the 
restframe B-band magnitude from a source the the limiting SNAP depth.}
\label{fig:color_omp3oxp7rv5p5_1}
\end{figure}

\begin{figure}
  \includegraphics[width=5.1cm]{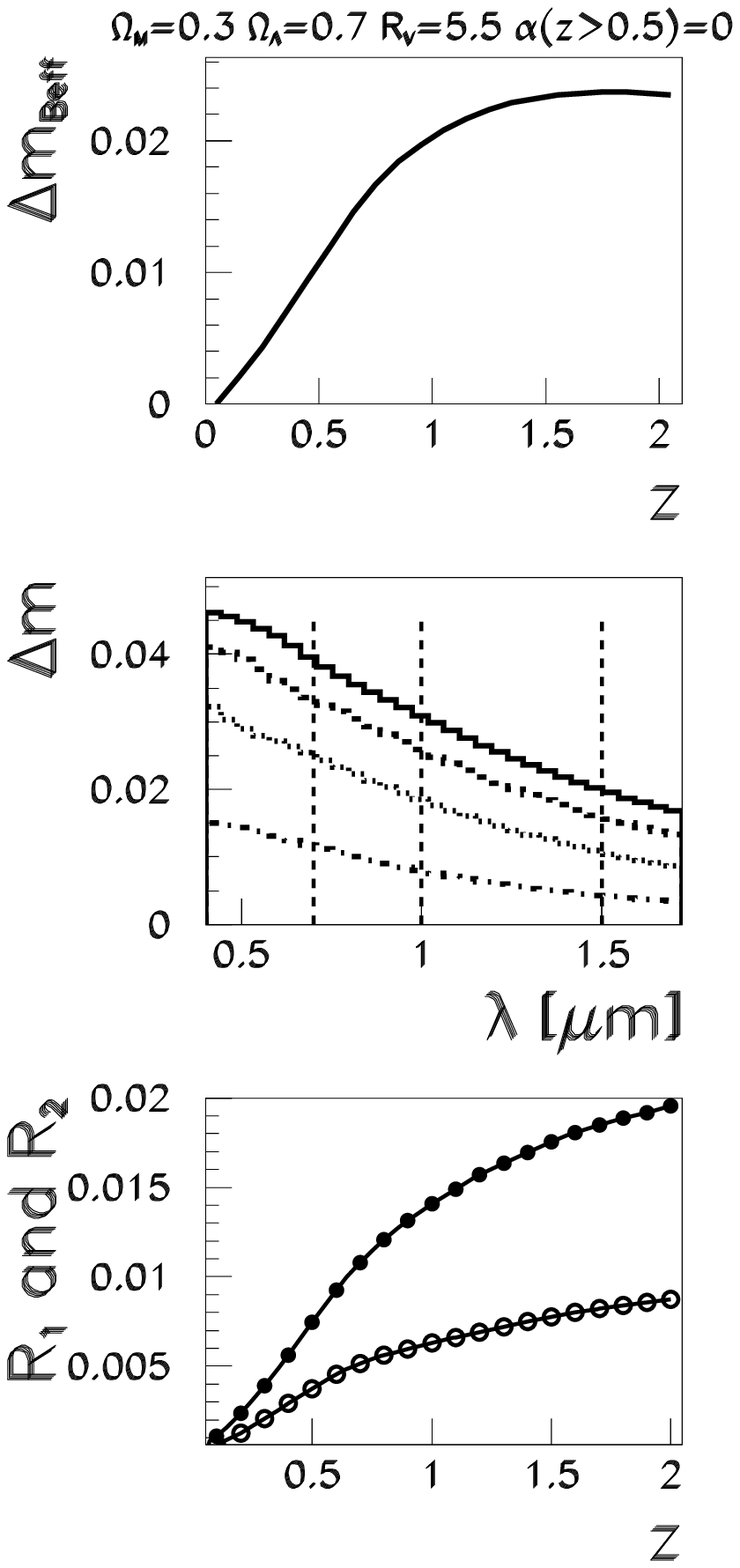}  \hspace*{-1.6cm}  
  \includegraphics[width=5.1cm]{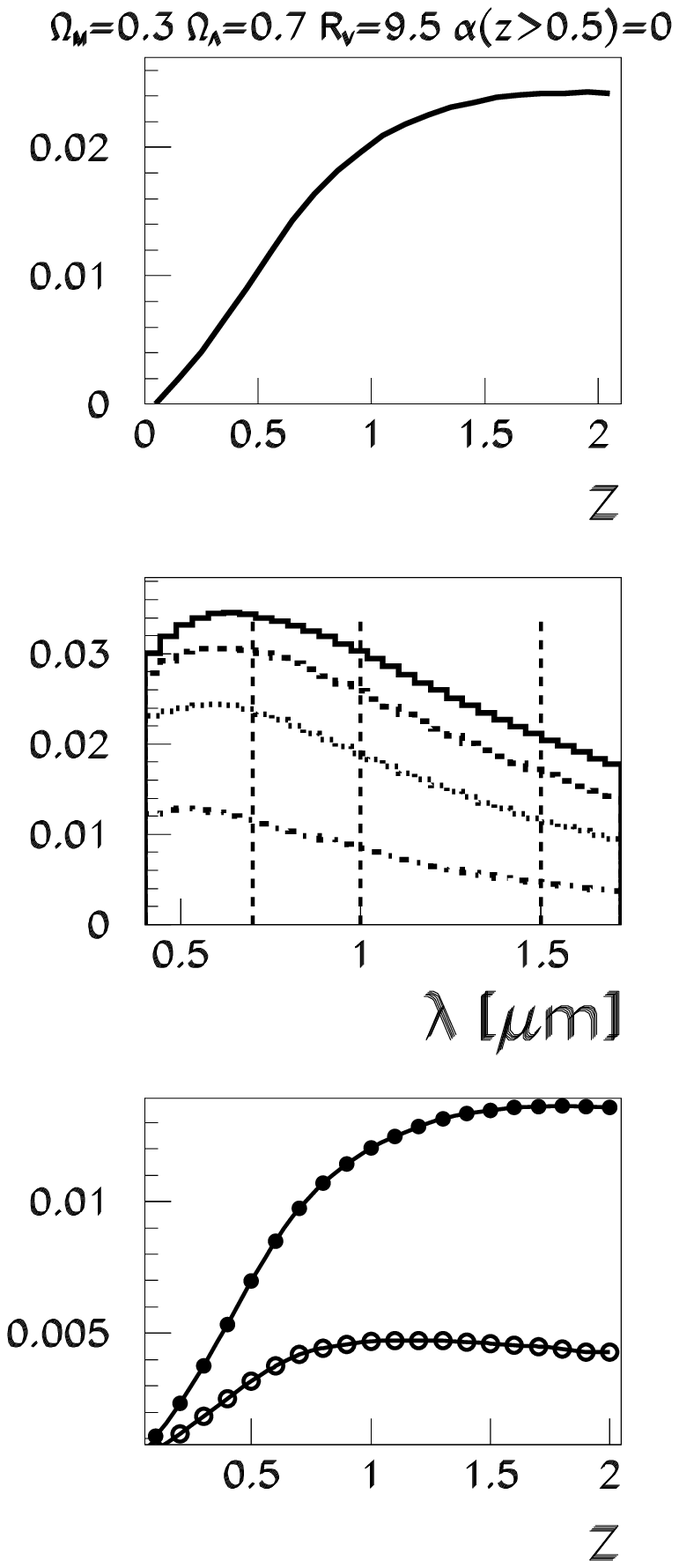}   
 \caption[]{Dust extinction in model B for $R_V$=5.5 (left side) and $R_V$=9.5 (right side) 
in a flat $\Lambda$-dominated universe with the dust density 
adjusted to generate about 0.02 mag dimming in the observed band corresponding to the 
restframe B-band magnitude from a source the the limiting SNAP depth
($\lambda_{\rm dust}$=300$\cdot\left({0.65 \over h}\right)$ Gpc).  
The upper panels show the dimming in the observational band corresponding to 
the restframe B-band. The middle panels show the differential extinction as a function of wavelength
for a source at redshifts $z=0.5,1.0,1.5$ and 2.0. The dashed lines show the position of $\lambda$=
0.7, 1.0 and 1.5 $\mu$m used to calculate $R_1$ and $R_2$. The bottom panels show the differential
color coefficients $R_1$ and $R_2$ as a function of redshift.}
  \label{fig:omp3oxp7rv9p5_2} 
\end{figure}

\begin{figure}
  \includegraphics[width=5.1cm]{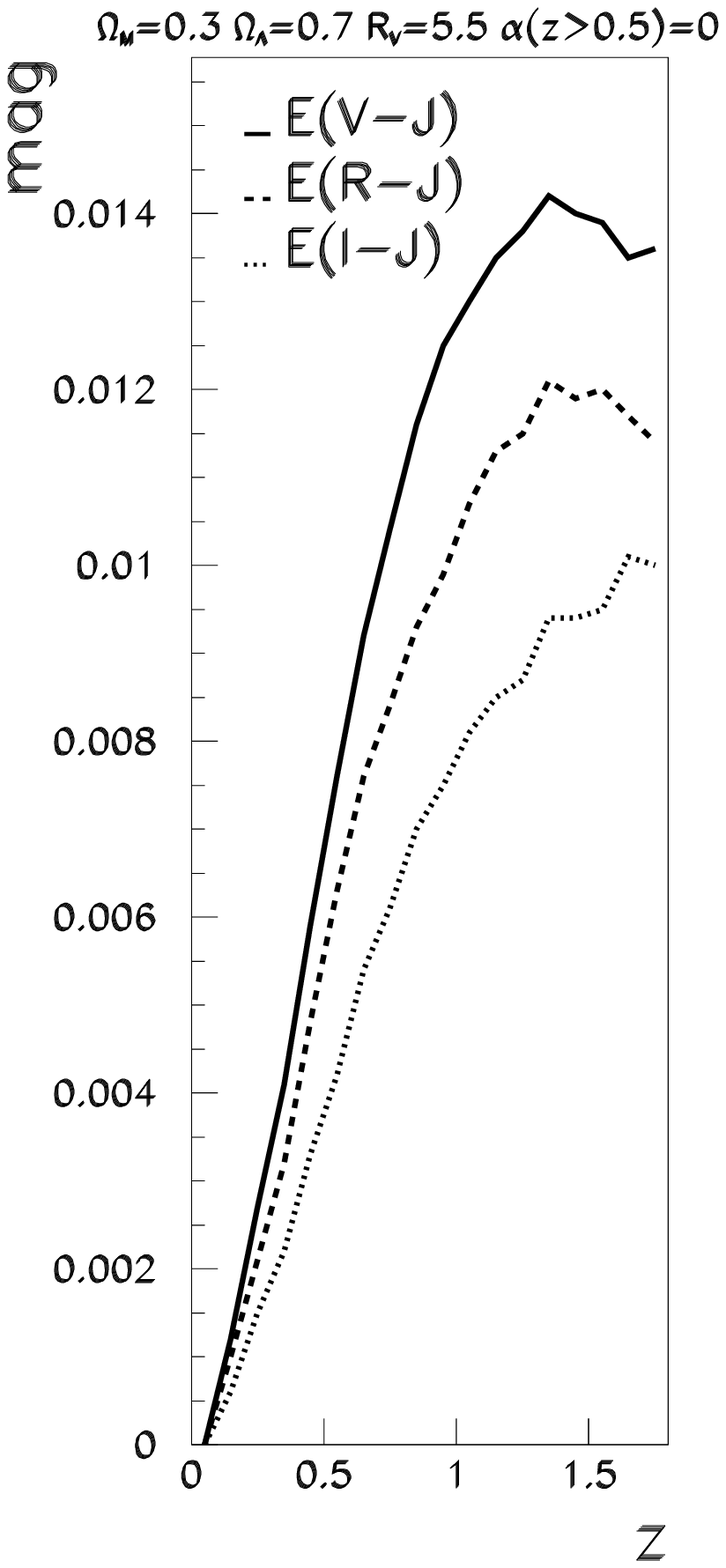}  \hspace*{-1.6cm}  
  \includegraphics[width=5.1cm]{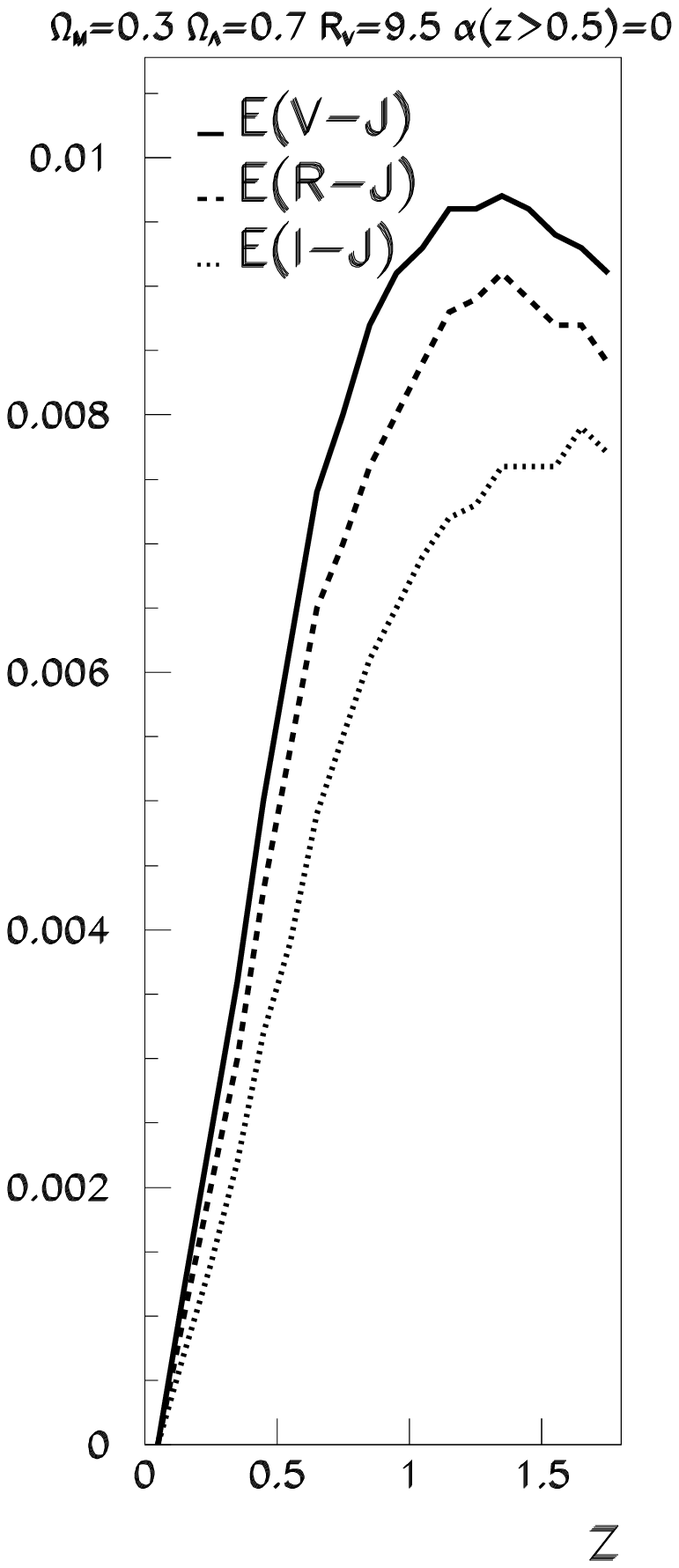}   
\caption[]{Color extinctions E(V-J),E(R-J) and E(I-J)  for Type Ia supernovae in dust model B 
 for $R_V$=5.5 (left side) and $R_V$=9.5 (right side) 
in a flat $\Lambda$-dominated universe with the dust density 
adjusted to generate about 0.02 mag dimming in the observed band corresponding to the 
restframe B-band magnitude from a source the the limiting SNAP depth.}
  \label{fig:color_omp3oxp7rv9p5_2} 
\end{figure}

Next, we consider the scenario where the extinction due to dust is negligible at
$z\sim 0.5$ but that it may introduce a bias in the Hubble diagram in excess of $\delta m$=0.02
for $z\le 2$. Thus, we consider a universe where $\Omega_{\rm M}$=0.3 and $\Omega_\Lambda=0.7$, e.g.  
Figs.\,~\ref{fig:omp3oxp7rv5p5_1} and \ref{fig:color_omp3oxp7rv5p5_1} show a case in dust model A.
At the limiting redshift range of SNAP a 0.1 magnitude
bias in the Hubble diagram results in a approximately a 6\% differential reddening for $R_2$
and 2\% in $R_1$ for $R_V$=5.5. For the same scenario but with $R_V$=9.5, $R_2$ exceeds 0.01 
for $z\gsim0.7$. The corresponding situation in 
model B can be seen in Figs.\,~\ref{fig:omp3oxp7rv9p5_2} and \ref{fig:color_omp3oxp7rv9p5_2}.

The ``greyness'' of $r>100$ nm dust grains make them completely 
undetectable with colors bluer than R-band in the restframe system. 
However, for longer wavelengths the extinction leaves a measurable signature in 
the observed V-J,R-J and I-J. This 
indicates that the {\em average} V-J, R-J and I-J colors of {\em normal} Type Ia supernovae
are significantly different with and without the extinction due to large 
dust grains. 

\section{Dust in galaxies along the line of sight}

As dust in galaxies is generally known to cause more reddening than the
hypothetical extragalactic ``grey'' component, intervening galaxies along
the line of sight cause less potential danger. Still, it may be of
interest to estimate the frequency and size of such encounters.  
In order to estimate the effects from dust due to intervening galaxies,
we have thus performed a number of ray-tracing Monte-Carlo simulations. 
We follow a light-ray between the source and the observer by sending the ray
through a series of spherical cells. Each cell encompasses a galaxy at the 
center and the size of the cell is computed to accurately represent the 
number density of galaxies, see (\cite{SNOC}) for a more detailed description
of the Monte-Carlo method. 

\begin{figure}
    \includegraphics[width=\hsize]{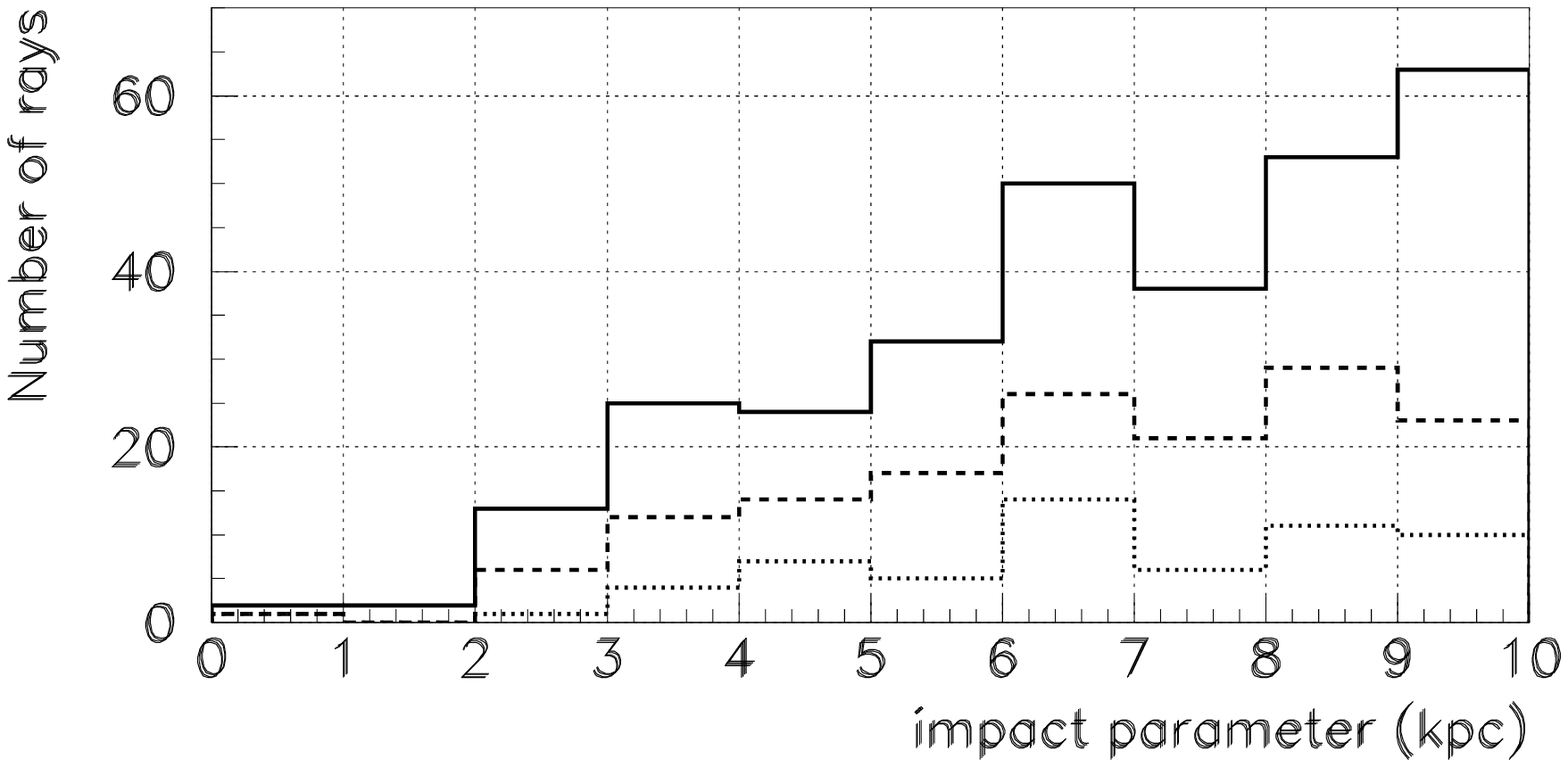} 
    \caption[]{The distribution of smallest galaxy impact parameters for 10\,000 sources at 
        redshifts $z=0.5$ (dotted line), $z=1$ (dashed line) and $z=1.5$
        (full line) in a $\Lambda$-dominated universe,  
        ($\Omega_{\rm M}$,$\Omega_\Lambda$)=(0.3,0.7).}
    \label{fig:zimpact}
\end{figure}

In Fig.~\ref{fig:zimpact} we have plotted the smallest galaxy impact parameters for 
10\,000 light-rays originating at redshifts $z=0.5, z=1$ and 
$z=1.5$ in a $\Lambda$-dominated universe, ($\Omega_{\rm M}$,$\Omega_\Lambda$)=(0.3,0.7). 
We see that the probability for a light-ray to pass closer than 10 kpc to 
a galaxy center is 0.6 \%, 1.5 \% and 3.0 \% for source redshifts of
$z=0.5, z=1$ and $z=1.5$ respectively. 

\begin{figure}[htp]
    \includegraphics[width=\hsize]{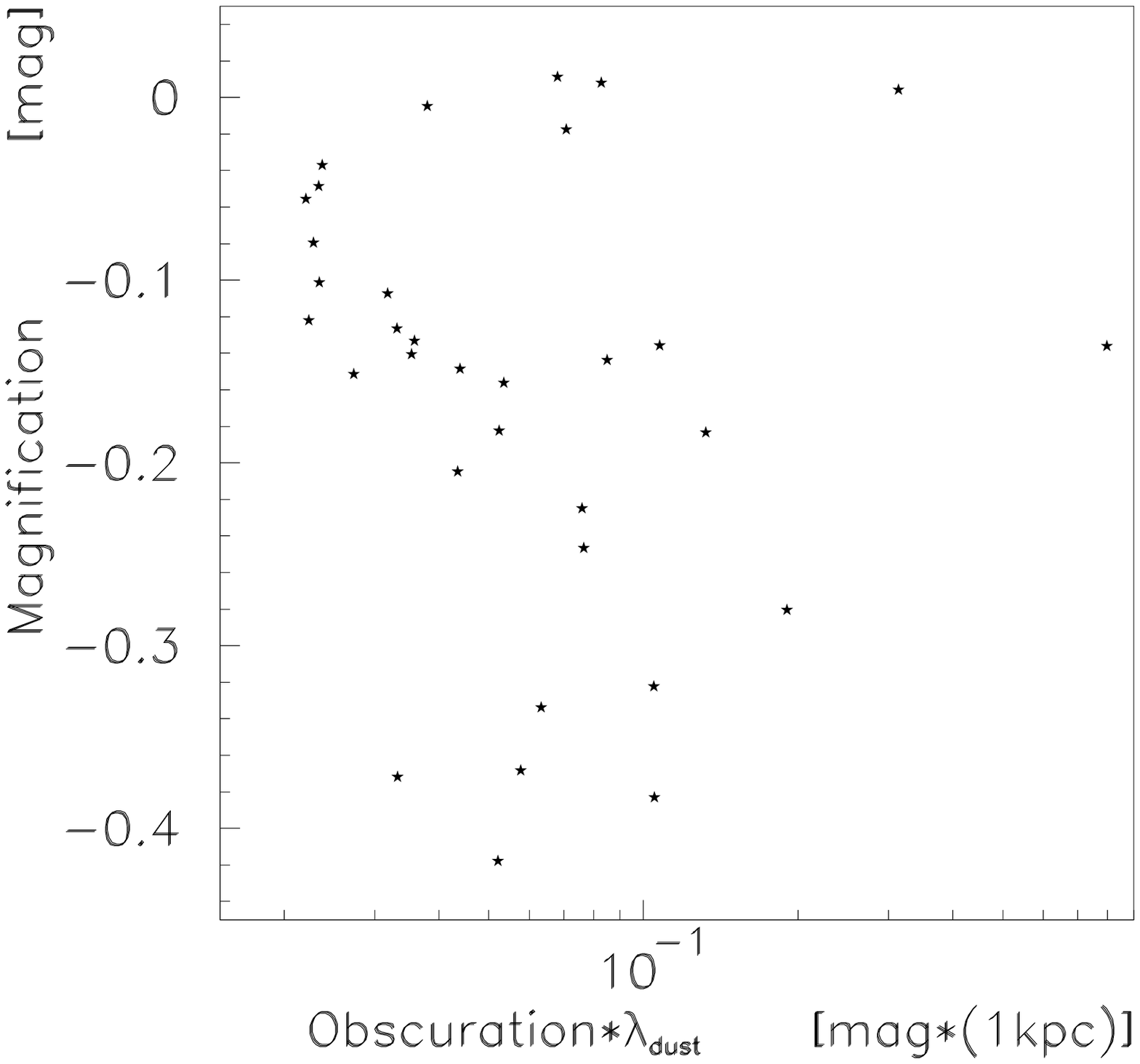} 
    \caption[]{The correlation of the demagnification due to dust 
        and the magnification due to gravitational lensing for the 33 out
        of 10\,000 sources at $z=1$ which are demagnified more than 
        0.02 mag by intervening galactic dust for an extinction
        scale-length at the galaxy centers
        of 1 kpc. To obtain the demagnification for an arbitrary scale-length,
        divide the numbers on the $x$-axis with the scale-length expressed in kpc.}
    \label{fig:glxydust}
\end{figure}

In order to estimate the magnitude of the effect, we model
the dust distribution in spiral galaxies by a double exponential disk with random
inclination,
\begin{equation}
  \label{disk}
  \rho_{\rm dust}=\rho^0_{\rm dust} e^{-r/r_0} e^{-\zeta/\zeta_0}.
\end{equation}
For simplicity, we assume that the dust-density scale-length in the 
$\zeta$-direction, $\zeta _0$, is constant
(0.1 kpc) whereas the radial scale-length, $r_0$, is proportional to the square-root of the 
luminosity of the galaxy, i.e., we assume that the distribution of dust  
follows the distribution of stars in the galaxy. For an $L_*$ galaxy, we use a 
scale-length $r_0=5$ kpc. The dust-disk is truncated at $\zeta=3$ kpc and $r=20$ kpc. 
We assume that the fraction of galaxies that are spirals is given by
\begin{equation}
  \label{nspiral}
  f_s(z)=f_s(0)+q\cdot z,
\end{equation}
where $f_s(0)=0.7$ and $q=0.05$. 
For an extinction scale-length at the galaxy centers of
1 kpc, we obtain a probability for a supernova at $z\sim 1$ 
to be demagnified by more than 0.02 mag by dust in intervening galaxies 
of $\sim 0.33$~\%. However, this number is very sensitive to the normalization of the
dust density since the absorption expressed in magnitudes is 
inversely proportional to the extinction scale-length of the absorption.
Since a light-ray passing close to a galaxy will be
magnified due to gravitational lensing effects, we expect some correlation 
of the demagnification due to dust and the magnification due to 
gravitational lensing. In Fig.~\ref{fig:glxydust}, the magnification from
lensing is plotted against the demagnification from dust in units of the inverse
extinction scale-length. To obtain the demagnification for a specific scale-length,
one should divide the numbers on the $x$-axis in Fig.~\ref{fig:glxydust} with the 
extinction scale-length expressed in kpc.

\section{Extinction in host galaxies}

\begin{figure}
    \includegraphics[width=\hsize]{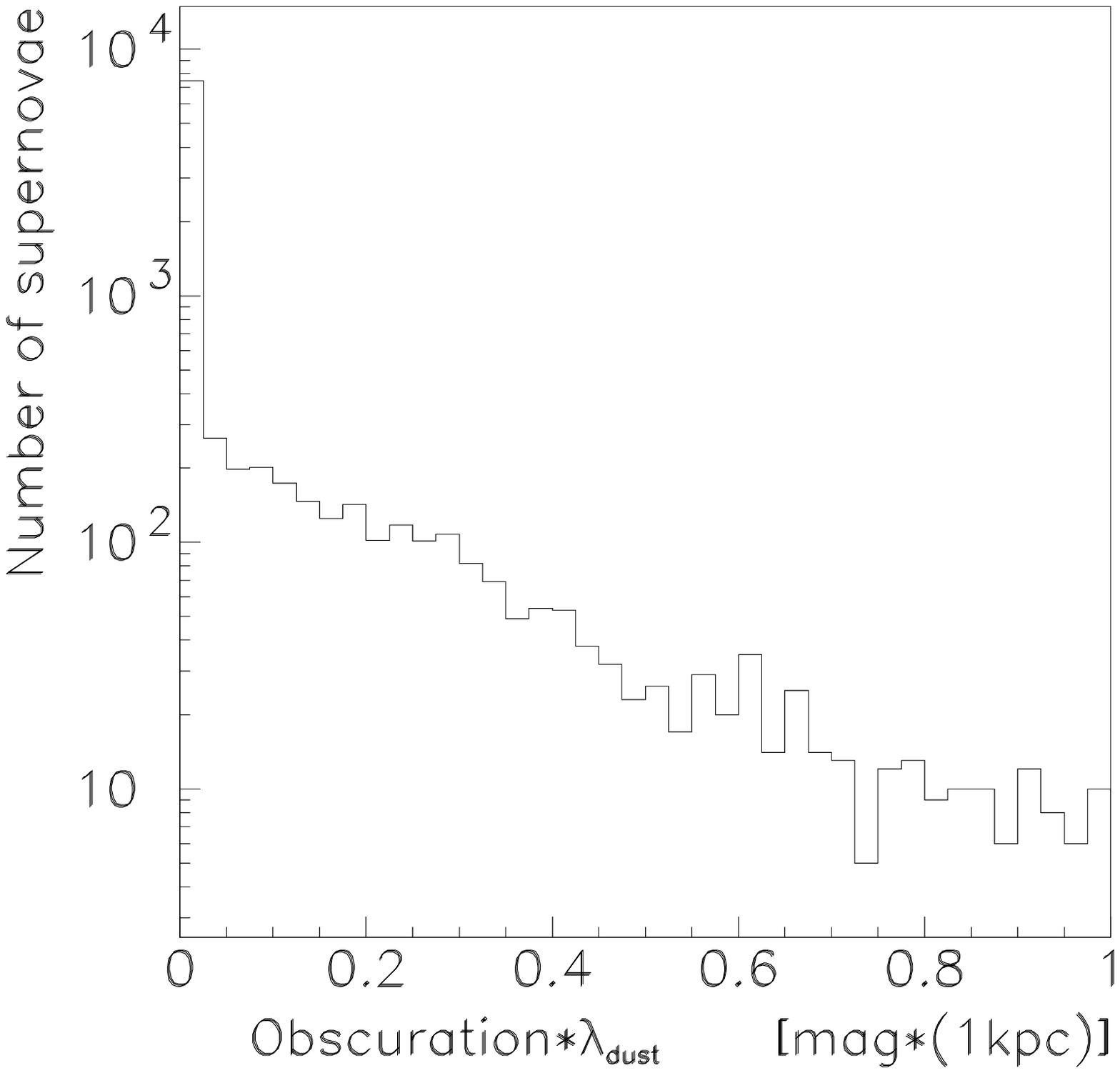} 
    \caption{The demagnification due to dust in the host galaxy.
        For an extinction scale-length of 1 kpc, $\sim$ 2600 out of 10\,000 sources 
        are demagnified by more than 
        0.02 mag by host galactic dust.}
    \label{fig:hostdust}
\end{figure}

Type Ia supernovae occur in both late and early type galaxies. In the case of
late type hosts, we need to take obscuration due to dust in the host galaxy
into account. We model the distribution of dust in late type host according to
Eq.~(\ref{disk}), and the generation of Type Ia SNe as two different components,
following (\cite{hatano}).
In the disk component, the generation of Type Ia SNe follows the dust distribution 
only that the scale-length in the $\zeta$-direction is increased to 0.35 kpc. There
is also a bulge component where the SNe are spherically distributed as
$(R^3+0.7^3)^{-1}$, where the distance from galactic center, $R$, 
is expressed in kpc and the bulge is truncated
at 3 kpc. We assume that the probability
for a SNe to occur in the bulge is 1/8 of the probability in the disk. 
Following (\cite{matteucci}), we estimate the probability for a Type Ia SNe to occur in an 
early type galaxy to be 
\begin{equation}
  \label{p}
  p_E(z)=p_E(0)+q\cdot z
\end{equation}
where $p_E(0)=0.5$ and $q=0.125$.
In Fig.~\ref{fig:hostdust}, we plot the obscuration due to 
host galaxy dust for 10\,000 Type Ia SNe. Note that the magnitude of the 
results is very sensitive to the normalization of the dust density. To obtain the
demagnification, divide the numbers on the $x$-axis with the extinction 
scale-length expressed in kpc. For an extinction scale-length of 1 kpc, 
$\sim$ 2600 out of 10\,000 sources are demagnified by more than 
0.02 mag by host galactic dust. 
%The rms-dispersion caused by this effect
%is 

%\begin{equation}
%  \label{sigma}
%  \sigma^{\rm host}_{\rm dust}\sim\frac{0.18}{\lambda_{\rm dust}^{\rm host}({\rm kpc})}\;{\rm mag}.
%\end{equation}
%As the observed dispersion of Type Ia supernovae does not, after shape calibration, 
%exceed 0.16 magnitudes, $\lambda_{\rm dust}^{\rm host} \gsim 1$~kpc.

\section{Discussion}

The effects of grey dust extinction capable of biasing the results of experiments such as
the proposed SNAP satellite can be diagnosed through accurate relative spectrophotometry or
broadband photometry at the 1\%
level in the $0.7-1.5$ $\mu$m wavelength range, as shown in e.g.  
Figs.\ \ref{fig:omp3oxp7rv9p5_2} and \ref{fig:color_omp3oxp7rv9p5_2}. This would allow tests
for intergalactic dust obscuration
affecting the measurement of high-z supernovae up to $\Delta m \sim 0.02$, the target for
systematic uncertainties for SNAP. 

Achieving 1-2\% relative spectrophotometric accuracy in the optical and
NIR for 22-29 magnitude objects is rather challenging. As the
measurements will rely on a large number of homogeneous high-redshift
sources, it is not required to get S/N=50-100 for individual
objects. On the other hand, the instrument calibration must be within
1\% over the course of at least one year for the case of SNAP.  This
can be accomplished e.g. through repeated observations of hot galactic
white dwarfs (\cite{bohlin}; \cite{finley}).
In the Rayleigh-Jeans limit, T$_{\rm eff}\gg$20000 K, the color of the calibrators
become independent of temperature and are therefore ideal for relative
calibration.

Significant progress in examining the possible bias due to grey dust in the 
published Type Ia supernova Hubble diagram can already be made with existing NIR 
instruments. 
If the faintness of Type Ia SNe at high-z is to be
attributed to grey dust obscuration as opposed to the cosmological
explanation a color extinction $E(R-J)^{\rm obs} \gsim 0.1$ mag is
to be expected for sources at $z\sim0.5$.
Testing this possibility is within reach with ground based
facilities.

We have also examined the possibility of extinction as the beam of
high redshift passes through foreground galaxies. This was found to be
a relatively small effect, only causing $>0.02$ mag dimming for less
than 1 \% of the sources at $z=1$, whereas extinction in the host
galaxy potentially is a more serious problem.
%This however, should lead
%to a broadening of the brightness dispersion of supernovae Ia. The relatively narrow measured
%spread of Type Ia supernovae indicates that the exponential dust scale length of the host galaxy 
%dust cannot be smaller than $\sim 1$ kpc.

However, in general we expect dust in galaxies along the line of sight and
in host galaxies to have $R_V\sim 3$, causing considerably more reddening than an
intergalactic "grey" dust component. Thus, with high accuracy spectrophotometry,
it should be possible to control the effects from extinction in galaxies.

\section*{Acknowledgements}
The authors would like to thank Anthony Aguirre, Alex Kim, and Peter Nugent for useful discussions and Serena Nobili for a careful reading of
the manuscript. 
AG is a Royal Swedish Academy 
Research Fellow supported by a 
grant from the Knut and Alice Wallenberg Foundation.
LB would like to thank the Swedish Research Council for financial support.


\begin{thebibliography}{}
% \bibitem[2001]{ramme}
%  Amanullah, R., et al. 2001, 
%  in preparation

\bibitem[1999a]{aguirre99a} A.~Aguirre, 1999, ApJ, 512, L19

\bibitem[1999b]{aguirre99b} A.~Aguirre, 1999, ApJ, 525, 583

\bibitem[Aguirre \& Haiman (2000) ]{aguirrehaiman}
 A. Aguirre and Z. Haiman, 2000, ApJ, 532, 28

\bibitem[Bahcall \& Fan 1998]{bahcall}N.A.~Bahcall and X.~Fan, 1998, ApJ, 504, 1

\bibitem[Balbi et al. 2000]{maxima}
A.~Balbi {\it et al.}, 2000, ApJ, 545,1; erratum: 2001, ApJ, 558, 145
\bibitem[Bohlin 1996]{bohlin}
R.C.~Bohlin, 1996, AJ, 111, 1743

\bibitem[Cardelli et al. 1989]{ccm}J.~A.~Cardelli, G.~C.~Clayton and J.~S.~Mathis,
1989, ApJ, 345, 245

\bibitem[de Bernardis et al. 2000]{boomerang}
P.~de Bernardis {\it et al.}, Nature 2000, 404, 955

\bibitem[Draine \& Lee 1984]{drainelee} B.~Draine and H.~Lee, 1984, ApJ, 285, 89

\bibitem[Finley et al. 1997]{finley} 
D.S~Finley, D.~Koester and G.~Basri, 1997, ApJ 488, 375

\bibitem[Goliath et al. 2001]{goliath} M.~Goliath, R.~Amanullah,
P.~Astier, A.~Goobar, and R.~Pain, 2001, A\&A in press, astro-ph/0104009

\bibitem[Goobar et al., 2001]{SNOC}  A.~Goobar et al., the SNOC Monte-Carlo package, in preparation, 2001

\bibitem[Hatano et al. 1998]{hatano}K.~Hatano et al. 1998, ApJ 502, 177

\bibitem[Ivanov et al. 2000]{ivanov}V.D~Ivanov et al. 2000, ApJ 542, 588

\bibitem[Madau 2000]{madau}P. Madau, Proc. of the Nobel Symposium ``Particle
Physics and the Universe'', L. Bergstr\"om, P. Carlson and 
C. Fransson (eds.), 2000, Physica Scripta, T85, 156

\bibitem[Matteucci \& Recchi 2001]{matteucci}F.~Matteucci and
S.~Recchi, 2001, ApJ, 558, 351

\bibitem[M\"ortsell et al. 2001]{mortsell}
E.~M\"ortsell, A.~Goobar and L.~Bergstr\"om, 2001, ApJ, 559, 53

\bibitem[Nugent et al. 2001]{nugent} P.~Nugent et al., 2001, submitted to PASP.

\bibitem[Peacock et al. 2001]{2DF} J.~A.~Peacock {\it et al.}, 2001, Nature, 410, 169 

\bibitem[Perlmutter et al. 1999]{scp}S.~Perlmutter et al., 1999, ApJ, 517, 565

\bibitem[Perlmutter et al. 2000]{snap} S.~Perlmutter et al., the SuperNova Acceleration Probe, http://snap.lbl.gov

\bibitem[Pryke et al. 2001]{dasi}
C.~Pryke {\it et al.}, 2001, submitted to ApJ, {\em pre-print} astro-ph/0104490 


\bibitem[Riess et al. 1998]{highz}A.~G.~Riess et al., 1998, AJ, 116, 1009 

\bibitem[Riess et al. 2000]{riessNIR} A.~Riess {\it et al.}, 2000, ApJ, 536, 62 

\bibitem[Riess et al. 2001]{riess17} A.~Riess {\it et al.}, 2001, ApJ, 560, 49

%\bibitem[Steidel ???]{steidel}

\bibitem[Steinhardt 2000]{steinhardt:haga}P.~J.~Steinhardt, 
Proc. of the Nobel Symposium ``Particle
Physics and the Universe'', L.~Bergstr\"om, P.~Carlson and 
C.~Fransson (eds.), 2000, Physica Scripta, T85, 177 


\bibitem[Vanden Berk et al. 2001]{SDSSquasars} D.~E.~Vanden Berk  {\it et al.}, 2001, astro-ph/010523 



%\bibitem[Netterfield et al., 2001]{boomerang}
%C.B.~Netterfield {\it et al.},{\em pre-print} astro-ph/0104460 (2001).

%\bibitem[Stompor et al., 2001]{maxima}
%R.~Stompor {\it et al.}, {\em pre-print} astro-ph/0105062 (2001).



%\bibitem[1997]{bowers}E.~J.~C.~Bowers {\it et al.}, MNRAS, 290, 663 (1997).
%\bibitem[Perlmutter et al. 2001]{snap} The SNAP proposal, snap.lbl.gov.
\end{thebibliography}
\end{document}